\documentclass[prb,a4paper,twocolumn,superscriptaddress]{revtex4}

\usepackage{amsmath,amssymb,dsfont}

\usepackage{graphicx}
\graphicspath{{./Figures/EPS/}}
\newcommand{\dgongwire}{(\Delta g/g)_\mathrm{wire}} 
\newcommand{\dgongdw}{(\Delta g/g)_\mathrm{dw}} 
\newcommand{\dgongint}{(\Delta g/g)_\mathrm{int}} 
\newcommand{\lsd}{l_\mathrm{sd}} 
\newcommand{\nimp}{n_\mathrm{i}} 
\newcommand{\Nimp}{N_\mathrm{i}} 

\newcommand{\zeromat}{\mathbf{0}} 
\newcommand{\idmat}{\mathbf{1}} 
\newcommand{\tmat}{t}%{\mathsf{t}} 
\newcommand{\rmat}{r}%{\mathsf{r}} 
\newcommand{\rmi}{\mathrm{i}} 
\newcommand{\rme}{\mathrm{e}}
 
\newcommand{\EF}{E_{\mathrm{F}}}
\newcommand{\pF}{p_{\mathrm{F}}}
\newcommand{\kF}{k_{\mathrm{F}}}

\newcommand{\eV}{\mathrm{eV}}

\newcommand{\nanometer}{\mathrm{nm}}

\renewcommand{\cos}{\mathop{\mathrm{cos}}}
\renewcommand{\sin}{\mathop{\mathrm{sin}}}

\renewcommand{\arccos}[1]{\mathrm{cos}^{-1}}
\renewcommand{\arcsin}[1]{\mathrm{sin}^{-1}}
\renewcommand{\arctan}[1]{\mathrm{tan}^{-1}}
\renewcommand{\exp}[1]{\mathrm{exp}\left(#1\right)}
 
\newcommand{\sgn}{\mathop{\mathrm{sgn}}}

\newcommand{\mymatrix}[4]{\begin{pmatrix}#1 & #2 \\ #3 & #4 \end{pmatrix}}
\newcommand{\spinor}[2]{\binom{#1}{#2}} 

\newcommand{\ket}[1]{\left|#1\right>}
\newcommand{\braket}[2]{\left<#1|#2\right>}

\newcommand{\var}{\mathrm{Var}}
\newcommand{\uni}{\mathrm{uni}}
\newcommand{\dw}{\mathrm{dw}}
\newcommand{\inc}{\mathrm{inc}}
\newcommand{\co}{\mathrm{co}}
\renewcommand{\eqref}[1]{Eq.~(\ref{#1})}
\newcommand{\eqsref}[1]{Eqs.~(\ref{#1})}

\newcommand{\ie}{\textit{i.e.\ }} 
\newcommand{\etal}{\textit{et al.\ }}
\newcommand{\stras}{\affiliation{Institut de Physique et Chimie des Mat\'eriaux
de Strasbourg, UMR 7504 (CNRS-ULP), 23 rue du Loess, Bo\^ite Postale 43, 67034
Strasbourg Cedex 2 FRANCE}} 
\newcommand{\uwa}{\affiliation{School of Physics, The University of Western
Australia, 35 Stirling Highway, Crawley WA 6009, AUSTRALIA}}

\begin{document}

%\ifpdf 
%\DeclareGraphicsExtensions{.pdf, .jpg, .tif} 
%\else
\DeclareGraphicsExtensions{.eps, .jpg} 
%\fi

\title{Electron Transport through Disordered Domain Walls: Coherent and
Incoherent Regimes} 
\author{Peter E.\ Falloon}
\uwa\stras 
\author{Rodolfo A.\ Jalabert} 
\email{jalabert@ipcms.u-strasbg.fr} 
\stras
\author{Dietmar Weinmann} 
\stras 
\author{Robert L.\ Stamps} 
\uwa

\begin{abstract} 
We study electron transport through a domain wall in a ferromagnetic nanowire
subject to spin-dependent scattering. A scattering matrix formalism is developed
to address both coherent and incoherent transport properties. The coherent case
corresponds to elastic scattering by static defects, which is dominant at low
temperatures, while the incoherent case provides a phenomenological description
of the inelastic scattering present in real physical systems at room
temperature. It is found that disorder scattering increases the amount of
spin-mixing of transmitted electrons, reducing the adiabaticity. This leads, in
the incoherent case, to a reduction of conductance through the domain wall as
compared to a uniformly magnetized region which is similar to the giant
magnetoresistance effect. In the coherent case, a reduction of weak
localization, together with a suppression of spin-reversing scattering
amplitudes, leads to an enhancement of conductance due to the domain wall in the
regime of strong disorder. The total effect of a domain wall on the conductance
of a nanowire is studied by incorporating the disordered regions on either side
of the wall. It is found that spin-dependent scattering in these regions
increases the domain wall magnetoconductance as compared to the effect found by
considering only the scattering inside the wall. This increase is most dramatic in
the narrow wall limit, but remains significant for wide walls.
\end{abstract}

\maketitle

\section{Introduction} 

The interplay between magnetic structure and electrical resistance in mesoscopic
ferromagnetic systems is interesting from both technological and fundamental
points of view. Magnetoresistance properties of domain walls represent a
particularly striking example of this rich physical problem, and have been
intensively studied in recent years. Experiments on ferromagnetic thin films
originally found that domain walls contribute to an enhancement of conductance,
\cite{ruediger1998, hong1998} although this was later understood to be due to
anisotropic magnetoresistance (AMR) effects. \cite{ruediger1999} Further
experiments on cylindrical Co nanowires \cite{ebels2000} and thin
polycrystalline Co films \cite{dumpich2002} have yielded a reduction in
conductance, distinct from the positive AMR contribution, which has been
attributed to scattering from magnetic domain walls.

From the theoretical side two main approaches exist to address this problem. On
one hand, first-principle calculations take into account realistic band
structures to study equilibrium properties. Assuming that the resulting one-body
wavefunctions are a good description of the many-body wavefunctions, a
calculational scheme to study electronic transport can be developed. According
to the details of the model, large domain wall magnetoresistances may be found.
\cite{vanhoof1999, kudrnovsky2001} On the other hand, phenomenological models
including only the essential features of the band structure are extremely useful
as they give insight into important physical mechanisms. The most widely used
approach is the $sd$ model, originally due to Mott, \cite{mott1936} which
separates the conduction ($s$) electrons from those responsible for the magnetic
structure ($d$). The exchange interaction between $s$ and $d$ electrons is
incorporated in the spin-splitting $\Delta$ of the conduction band. An important
advantage of this model is that it can be readily extended to include impurity
scattering through a disorder potential, making it particularly suitable for the
topic of the present work.

The simplest approach is to treat transport through the domain wall and the
leads as ballistic, where the only scattering is due to the rotating
spin-dependent potential in the wall. The resulting magnetoconductance is very
small for typical domain wall widths, \cite{cabrera1974, weinmann2001} although
it becomes significant in the special case of domain walls trapped in magnetic
nanocontacts. \cite{imamura2000, nakanishi2000, dugaev2003} In general, the most
important effect of a domain wall on ballistic electron transport is a mixing of
the up and down spin channels, which arises from the inability of electrons to
follow adiabatically the local magnetization direction, referred to as
``mistracking''. \cite{viret1996, gopar2004} When spin-dependent scattering in
the regions adjacent to the domain wall is taken into account,\cite{falloon2004}
this mistracking of spin in a ballistic wall leads to a significant
magnetoconductance analogous to the giant magnetoresistance effect.
\cite{valet1993} 

However, treating the domain wall as a completely ballistic (disorder-free)
system is not realistic for experimentally relevant systems. For instance, the
cobalt nanowires of Ref.~\onlinecite{ebels2000} have an estimated elastic mean
free path of $\sim$$7\nanometer$, while the wall width in cobalt is
$\sim$$15\nanometer$. Since these two characteristic length scales are of the
same order, the transport through the domain wall cannot be described as either
ballistic or diffusive. For other materials, such as iron or nickel, the wall
widths are larger and, depending on the amount of disorder, the diffusive regime
may be reached. 

A number of works have focused on the role of disorder scattering inside the
domain wall. Viret \etal \cite{viret1996} used the ballistic mistracking of spin
mentioned above to develop an intuitive picture based on a weighted average of
up and down resistances. The estimated relative magnetoresistance decreases with
the width of the domain wall and agrees with measurements on domain wall arrays
in thin films. Similar results were found using models based on the Boltzmann
equation \cite{levy1997} and the Kubo formula.\cite{brataas1999} Phase coherence
effects have also been studied, \cite{tatara1997,lyandageller1998} and in the
case of spin-independent disorder a negative contribution to domain wall
magnetoconductance has been predicted. \cite{tatara1997,jonkers1999}

In this work we develop a new model for transport through a disordered domain
wall based on combining scattering matrices for individual impurity scatterers,
which improves on existing treatments in several key aspects. Firstly, the model
is non-perturbative in wall width and disorder strength, which allows us to
study walls of arbitrary width and consider both ballistic and diffusive
transport regimes. It is found that impurity scattering inside the domain wall
causes an increase in transmission and reflection with spin-mistracking, or
equivalently a reduction in the adiabaticity of spin transport through the wall.
Secondly, we can treat both phase-coherent and incoherent transport regimes
within the same model. This permits a quantitative determination of the
contribution of phase coherence effects to domain wall magnetoconductance. In
this way we find that for incoherent transport a domain wall gives rise to a
positive magnetoconductance effect (\ie a reduction of conductance), which
depends on the relative impurity scattering strength for up and down electrons
and scales linearly with the number of conductance channels. In constrast, for
coherent transport in the case of strong disorder, the domain wall
magnetoconductance is negative and does not depend on the relative up/down
scattering strength or the system size. Finally, spin-dependent scattering in
the regions adjacent to the domain wall can be incorporated directly into the
model, allowing us to calculate the total magnetoconductance effect of a domain
wall in a nanowire using an approach similar to the circuit model developed in
Ref.~\onlinecite{falloon2004}. It is found that scattering in the uniformly
magnetized regions on either side of the wall causes an enhancement of the
magnetoconductance effect which is largest for narrow walls but is also
significant in the wide wall limit.

The layout of this paper is as follows. In Section II we introduce our physical
model and describe the numerical method which we use to calculate conductance
through a disordered region. In Section III we study the intrinsic transport
properties of a disordered domain wall in both coherent and incoherent regimes.
In Section IV we incorporate spin-dependent scattering in the regions adjacent
to the wall. Finally, in Section V we discuss the experimental relevance of our
findings, presenting our conclusions and outlook.

%%%%%%%%%%%%%%%%%%%%%%%%%%%%%%%%%%%%%%%%%%%%%%%
\section{Physical model} \label{sect:physical model}

We consider a quasi one-dimensional (quasi-1D) wire, with longitudinal axis
lying along the $z$ axis and a single transverse dimension of length $L_y$ (see
Figure \ref{fig:disordered wall}). The position $\vec{r}$ denotes the
two-dimensional vector $(y,z)$. The extension to a two-dimensional cross section
is straightforward, but complicates the notation and numerical calculations
without adding new physics. Within the $sd$ model, the conduction electrons in
the wire are described by a free-electron one-body hamiltonian with
spin-dependent potential
\begin{equation}\label{eq:hamiltonian}
H = -\frac{\hbar^2}{2m}\nabla^2 + 
\frac{\Delta}{2}\vec{f}(\vec{r})\cdot\vec{\sigma}
 + V(\vec{r})\, .
\end{equation}
Here $\vec{f}(\vec{r})$ is a unit vector in the direction of the effective field
representing the magnetic moment due to the $d$ electrons, $\vec{\sigma}$ is the
Pauli spin operator of the $s$ electrons, and $\Delta$ is the spin-split energy
gap between up and down $s$ electrons. The potential $V(\vec{r})$ represents
impurities and leads to scattering. It is discussed in detail in Section
\ref{sub:delta function model}.

For a wire uniformly magnetized in the $z$ direction we have $\vec{f}(\vec{r}) =
(0,0,-1)$ for all $\vec{r}$. When the wire contains a domain wall separating
regions of opposite magnetization along the $z$ axis, we have $\vec{f}(\vec{r})
= (0,0,-1)$ for $z<0$ and $\vec{f}(\vec{r}) = (0,0,1)$ for $z>\lambda$, with the
length $\lambda$ defining the wall region $0\le z\le\lambda$. 

For simplicity we assume a square well potential with infinite walls at
$y=0,L_y$ for the transverse confinement. In the disorder-free regions of
constant magnetization at either end of the system (\ie the leads), electrons
then occupy well-defined modes (channels):
\begin{equation}\label{eq:transverse phi}
\phi_n(y) = \sqrt{\frac{2}{L_y}} \sin\left(\frac{n\pi y}{L_y}\right),
\end{equation}
where $n$ is a positive integer.

The eigenstates in the leads (which constitute the asymptotic states in a
scattering approach to the domain wall) are characterized by a longitudinal
wavevector $k$, transverse mode number $n$, and spin eigenvalue $\sigma=\pm$
(representing spin states which are, respectively, anti-parallel and parallel to
the local direction of $\vec{f}(\vec{r})$). The corresponding dispersion
relation is
\begin{equation} \label{eq:E dispersion relation} 
E = \frac{\hbar^2 k^2}{2m}  
  + \frac{\hbar^2}{2m}\left(\frac{n\pi}{L_y}\right)^2 
  - \frac{\sigma\Delta}{2}.
\end{equation}
The relevant states for transport are those with energy $E$ at the Fermi energy
$\EF$, and corresponding wavevectors
\begin{equation}
k_{\sigma n} = \sqrt{k_{\sigma,\mathrm{F}}^2 - \left(\frac{n\pi}{L_y}\right)^2}.
\end{equation}
Here $k_{\sigma,\mathrm{F}} = \sqrt{\kF^2 + \sigma k_\Delta^2/2}$ is the
spin-dependent Fermi wavevector, defined in terms of $\kF=\sqrt{2m\EF/\hbar^2}$
and $k_\Delta = \sqrt{2m \Delta /\hbar^2}$. The number of propagating modes in
each lead and for each spin sub-band, $N_\sigma$, is given by the largest value
of $n$ for which $k_{\sigma n}$ is real ($N_\sigma \simeq k_{\sigma, \mathrm{F}}
L_y/\pi$).

The wavefunctions are two-component spinors:
\begin{equation}\label{eq:plain basis states}
\psi_{\sigma n}^\gtrless(\vec{r}) = 
\sqrt{\frac{\hbar}{m v_{\sigma n}}}
\rme^{\pm\rmi k_{\sigma n} z}\phi_n(y)
\ket{\sigma},
\end{equation}
where $\gtrless$ denotes propagation to the right/left, and the spinor basis
states $\ket{\pm}$ represent spin eigenstates parallel or anti-parallel to the
local value of $\vec{f}(\vec{r})$. The states are normalized to unit flux, where
the velocity factor for a plane-wave state is $v_{\sigma n} = \hbar
k_{\sigma n}/m$. 

Defining the spin states $\ket{\sigma}$ with respect to the local magnetization
direction allows us to treat both of the leads (as well as the uniform case)
within the same notation. Inside the wall ($0\le z\le \lambda$), $\ket{\sigma}$
depend on $\vec{r}$, and hence the relevant electron eigenstates are more
complicated than the states in \eqref{eq:plain basis states}. They are described
in Section \ref{sub:ballistic wall states}.

The parameters $\EF$ and $\Delta$ characterize the structure of the parabolic
$s$ band in the $sd$ model, and must be chosen to approximate the spin
polarization obtained from band structure calculations. This approximation
introduces considerable uncertainty. In this work we find that the dependence on
$\EF$ and $\Delta$, as well as the wall width $\lambda$, can be encapsulated in
a dimensionless ``effective'' wall width $\pF$ (\eqref{eq:pF def}). Throughout
this work, we fix $\EF=10\eV$ and $\Delta=0.1\eV$, which correspond to
reasonable parameter estimates for cobalt. \cite{getzlaff1996} By varying
$\lambda$, we then obtain results which correspond to different effective domain
wall widths and apply to arbitrary materials. 

\subsection{Scattering matrix approach to conductance}
\label{sub:S matrix approach} 

Since we assume non-interacting electrons, the conductance $g$ (in units of
$e^2/h$) may be described by the Landauer-B\"uttiker formula
\cite{buttiker1985}:
\begin{equation}
g = \sum_{\sigma',\sigma=\pm} 
\left\{
\sum_{n'=1}^{N_{\sigma'}} \sum_{n=1}^{N_\sigma}
T_{\sigma' n';\sigma n}(\EF)
\right\},
\end{equation}
where $T_{\sigma'n';\sigma n}(\EF)$ is the probability of an electron at energy
$\EF$ to be transmitted from the state with spin and transverse mode
$(\sigma,n)$ in the left lead to the one with $(\sigma',n')$ in the right lead.

\begin{figure}
\includegraphics[width=0.45\textwidth]{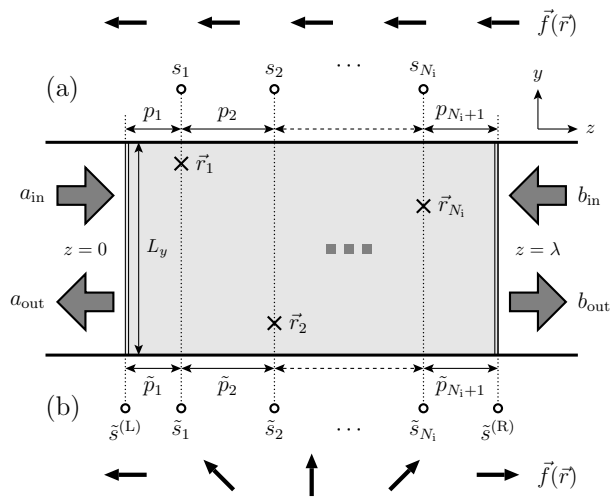}
\caption{\label{fig:disordered wall}
Schematic illustration of the scattering processes occurring in our model of a
disordered wire with (a) uniform magnetization and (b) a domain wall. Above and
below the wire are arrows indicating the magnetization direction,
$\vec{f}(\vec{r})$, as well as the scattering matrices describing the transport
through the wire in each case. The impurities are represented by point
scatterers randomly located at $\vec{r}_\alpha$ ($\alpha=1,\ldots,\Nimp$), with
corresponding scattering matrices $s_\alpha$ (uniform) and $\tilde{s}_\alpha$
(domain wall). The matrices $p_\alpha$ and $\tilde{p}_\alpha$ represent the
ballistic propagation between successive scatterers, while
$\tilde{s}^{(\mathrm{L,R})}$ represent the scattering from the domain wall
interfaces (case (b) only). These matrices can be combined in a coherent or
incoherent way to give the total scattering matrix for the system
(\eqsref{eq:s comb coherent} and (\ref{eq:s comb incoherent}) respectively). The
incident and scattered fluxes, $a_\mathrm{in}$, $b_\mathrm{in}$ and
$a_\mathrm{out}$, $b_\mathrm{out}$, are also shown.}
\end{figure}

To calculate the transmission amplitudes and probabilities, we use an approach
based on combining the scattering matrices of all the scatterers in the system,
which was developed to treat transport in spin-independent disordered systems by
Cahay \etal \cite{cahay1988} 

Our composition of scattering matrices neglects the effect of evanescent modes,
which have imaginary wavevector and are assumed to decay between successive
scattering events. This approximation has been widely used in the so-called
local approach for the random matrix description of quasi-1D wires.
\cite{beenakker1997, mello1992, stone1991} The impressive success of random
matrix theory in explaining the universal features of quantum transport, as well
as its agreement with numerical simulations and microscopic theories, relies on
the hypothesis of a quasi-1D geometry ($L_z \gg L_y$) and weak scattering ($l
\gg \delta L \gg \kF^{-1}$, where $\delta L$ is the size of the scattering
blocks). In our case these two hypotheses are valid, since the mean distance
between scatterers is much larger than the Fermi wavelength.

For the problem of spin-dependent transport our method offers several important
advantages over the spin-dependent extension of the recursive Green's function
(RGF) technique. The latter calculates the phase-coherent conductance with a
tight-binding approximation for \eqref{eq:hamiltonian} in which the rotating
magnetization and the impurity potential are incorporated through spin-dependent
on-site potential energies. \cite{frustaglia2001, gopar2004} In particular, the
scattering matrix approach allows the conductance to be calculated either
coherently or incoherently, which is useful for identifying and understanding
phenomena arising from phase coherence. Additionally, because scattering
matrices are expressed in terms of local basis states, we are able to avoid the
coarse-grained discretization inherent in the tight-binding description of the
domain wall magnetization. Finally, because only propagating states are included
in the scattering matrices, the computational effort required for a system with
a given number of conducting channels is significantly lower than with the RGF
technique. In the case of coherent transport through a disordered domain wall,
we have verified that results using the scattering matrix approach are in
quantitative agreement with those obtained using the RGF technique.
\footnote{P.~E. Falloon, PhD thesis, University of Western Australia -
Universit\'e Louis Pasteur (unpublished).}
 
The scattering matrix of a system relates incoming ($a_\mathrm{in}$,
$b_\mathrm{in}$) and outgoing ($a_\mathrm{out}$, $b_\mathrm{out}$) flux
amplitudes from the left and right (Figure \ref{fig:disordered wall}) through
\begin{equation}
\spinor{b_\mathrm{out}}{a_\mathrm{out}} 
= s \cdot \spinor{a_\mathrm{in}}{b_\mathrm{in}} 
= \mymatrix{\tmat}{\rmat'}{\rmat}{\tmat'} \cdot 
  \spinor{a_\mathrm{in}}{b_\mathrm{in}}.
\end{equation}
Here $\tmat,\rmat$ ($\tmat',\rmat'$) are transmission and reflection matrices
for states incident from the left (right). The elements of these sub-matrices
are the scattering amplitudes between individual modes $(\sigma,n)$ and
$(\sigma',n')$, which we write generically as $\xi_{\sigma'n';\sigma n}$, where
$\xi$ stands for one of $t$, $r$, $t'$, or $r'$. The dimension of these
sub-matrices of $s$ is determined by the number of propagating modes of each
spin on either side of the scatterer.

Throughout this work we will denote amplitudes and their matrices by lower-case
letters and the corresponding probabilities (given by the absolute square of the
amplitudes) by upper-case letters. We therefore use $s$ to denote scattering
matrices containing amplitudes and $S$ to denote the corresponding matrix of
probabilities. This convention differs from the standard notation in which $S$
denotes the scattering matrix of amplitudes, but we adopt it here for notational
consistency. The sub-matrices of $S$ are the transmission and reflection
probability matrices $T$, $R$, $T'$, and $R'$. We refer to these matrices
generically as $\Xi$, and the corresponding individual probabilities as 
$\Xi_{\sigma'n';\sigma n} = \lvert\xi_{\sigma'n';\sigma n}\rvert^2$.
We will also make use of the following notation for sums of transmission and
reflection probabilities summed over transverse modes and spin:
\begin{subequations}
\label{eq:summed tr probabilities}
\begin{eqnarray} 
\label{eq:total spin-dep tr}
\Xi_{\sigma'\sigma} &=& \sum_{n'n}\Xi_{\sigma'n';\sigma n}, 
\label{eq:total sd tr}\\
\Xi_\sigma &=& \sum_{\sigma'}\Xi_{\sigma'\sigma}.
\label{eq:total tr of incident spin}
\end{eqnarray}  
\end{subequations}
Here $\Xi_{\sigma'\sigma}$ represents the total scattering probability from
spin sub-band $\sigma$ to $\sigma'$, and $\Xi_\sigma$  the total scattering
probability (into both spin sub-bands) for states incident with spin $\sigma$.

For two scatterers described by the individual scattering matrices $s_1$ and
$s_2$, the resultant scattering matrix obtained by combining them in series is
written as $s_{12} = s_1 \ast s_2$, where $\ast$ stands for the composition law
\begin{subequations} \label{eq:S matrix combination}
\begin{eqnarray}
\tmat_{12} &=& \tmat_2\cdot(\idmat - \rmat'_1\cdot\rmat_2)^{-1}\cdot\tmat_1 ,\\
\rmat_{12} &=& \rmat_1 + \tmat'_1\cdot\rmat_2 (\idmat - \rmat'_1\cdot\rmat_2)^{-1}
\cdot\tmat_1, \\
\tmat'_{12} &=& \tmat'_1 (\idmat + \rmat_2 (\idmat -\rmat'_1\cdot\rmat_2)^{-1}
\cdot\rmat'_1)\cdot\tmat'_2, \\
\rmat'_{12} &=& \rmat'_2 + \tmat_2\cdot(\idmat - \rmat'_1\cdot\rmat_2)^{-1}
\cdot\rmat'_1\cdot\tmat'_2.
\end{eqnarray}
\end{subequations}
Here $\idmat$ denotes the identity matrix with the same number of rows as
$\rmat'_1$ and the same number of columns as $\rmat_2$. The application of this
composition law to include many scatterers forms the basis of the approach used
in this work.

\subsubsection{Coherent and incoherent transport through a disordered region}

In this work we are interested in the conductance through a disordered region
with either a domain wall or uniform magnetization. As we shall discuss in
Section \ref{sub:delta function model}, the effect of disorder is represented in
our model by the potential $V(\vec{r})$, which is comprised of $\Nimp$ delta
function scatterers at randomly distributed positions $\vec{r}_\alpha$. The
indices are ordered so that $z_\alpha<z_{\alpha+1}$ for $\alpha=1,\dots,\Nimp$.

Within the scattering matrix approach, the phase-coherent transmission from
$z=0$ to $\lambda$ is calculated by combining the scattering matrices of each
delta function scatterer with those for ballistic propagation between the
scatterers. We denote these matrices respectively by $s_\alpha$, $p_\alpha$ for
uniform magnetization, and $\tilde{s}_\alpha$, $\tilde{p}_\alpha$ for a domain
wall. In the case of a domain wall, we also require scattering matrices
$\tilde{s}^{(\mathrm{L,R})}$ to describe the scattering at the interfaces
between the domain wall and uniform regions. These matrices, and the basis
states inside the domain wall, are discussed in detail in the next sub-section
and in Appendix \ref{app:dw S matrix}. 

The matrices $p_\alpha$ and $\tilde{p}_\alpha$ contain the phase shifts acquired
by electrons propagating ballistically between scatterers at
$\vec{r}_{\alpha-1}$ and $\vec{r}_\alpha$. We write the longitudinal propagation
distances as $\delta z_\alpha = z_\alpha - z_{\alpha-1}$, where $z_1,\dots
z_{\Nimp}$ are the longitudinal components of the impurity positions and we
define $z_0=0$, $z_{\Nimp+1}=\lambda$. For the uniform case we have
\begin{equation} \label{eq:plain P}
p_\alpha = \mymatrix{\varphi_\alpha}{\zeromat}{\zeromat}{\varphi_\alpha},
\end{equation}
where 
\begin{equation}
\left[\varphi_\alpha\right]_{\sigma'n';\sigma n} =
\delta_{\sigma'\sigma} \delta_{n'n}
\exp{\rmi k_{\sigma n} \delta z_\alpha}.
\end{equation}
The matrices $\tilde{p}_\alpha$ for the domain wall case are defined in
exactly the same way within the basis of domain wall functions (defined in
\eqref{eq:spiral basis state}), with the domain wall wavevectors
$\tilde{k}_{\sigma n}$ (\eqref{eq:k spiral}) in place of $k_{\sigma n}$.

The total scattering matrices for coherent propagation through the disordered
region in the case of a domain wall and uniform magnetization can be
written respectively as
\begin{subequations}\label{eq:s comb coherent}
\begin{eqnarray}
s^{(\dw)}_\co &=&  
\tilde{s}^{(\mathrm{L})} \ast \tilde{p}_1 \ast \tilde{s}_1 \ast \tilde{p}_2
\ast \cdots \ast \tilde{s}_{\Nimp} \ast \tilde{p}_{\Nimp+1} \ast
\tilde{s}^{(\mathrm{R})}, \label{eq:sdw comb coherent} \nonumber \\ \\
s^{(\uni)}_\co &=&
p_1 \ast s_1 \ast p_2 \ast \cdots \ast s_{\Nimp} \ast p_{\Nimp+1}.
\label{eq:sndw comb coherent}
\end{eqnarray}
\end{subequations}

For the incoherent conductance, the phase coherence between successive
scattering events is assumed to be lost. The resulting scattering matrices for
the domain wall and uniform cases are obtained from the combination law of
\eqref{eq:S matrix combination}, but using probabilities ($\Xi$) instead of
amplitudes ($\xi$). We then write
\begin{subequations}\label{eq:s comb incoherent} 
\begin{eqnarray}
S^{(\dw)}_\inc &=& \tilde{S}^{(\mathrm{L})}\ast\tilde{S}_1\ast
\tilde{S}_2\ast\cdots\ast\tilde{S}_{\Nimp}\ast\tilde{S}^{(\mathrm{R})}, 
\label{eq:sdw comb incoherent} \\ 
S^{(\uni)}_\inc &=& S_1\ast S_2 \ast \cdots \ast
S_{\Nimp}. 
\label{eq:sndw comb incoherent} 
\end{eqnarray} 
\end{subequations}
The matrices $p_\alpha$ and $\tilde{p}_\alpha$ do not appear in this case since
they contain a pure phase shift and the associated transmission probabilities
are simply unity.

In the quasi-ballistic regime, the elastic mean free path is much larger than
the system size and only first-order scattering processes are important. For the
uniformly magnetized case the incoherent and coherent conductances become equal
in this regime. In the diffusive regime, on the other hand, constructive
interference of time-reversed paths with identical starting and end points leads
to an enhancement of the coherent reflection known as weak localization.
\cite{bergmann1984} In the limit of large disorder, the coherent conductance
enters the strongly localized regime, where $g$ decreases exponentially with
length and the system becomes insulating. \cite{kramer1993} By constrast, the
incoherent conductance scales like the classical Drude conductance ($g\sim
1/L$), even for arbitrarily large disorder. 

A characteristic feature of phase coherence effects in the diffusive regime
(such as weak localization or conductance fluctuations) is their universality,
or independence of system size, which means that their relative importance is
largest for small system sizes. In our model this is particularly significant
since we are constrained to work with system sizes that are smaller than typical
systems on which experiments are performed. Therefore, the relative importance
of coherence effects appears overemphasized. On the other hand, the incoherent
conductance scales with the system size and our numerical results can be safely
extrapolated to larger sizes. Furthermore, in the incoherent case the
fluctuations of conductance decrease with increasing disorder, which allows
average quantities to be computed accurately with relatively small numbers of
samples.

A fundamental limitation with finite numerical models of coherent transport is
that, for a given width $L_y$ and impurity density, there is a maximum value of
the system length $L_z$ beyond which the system enters the strongly localized
regime. However, if $L_z$ is larger than the phase coherence length $L_\phi$
then the localized behaviour obtained in the coherent model is not relevant. A
comparison between coherent and incoherent results permits us to determine those
features which are due to coherence and will disappear with increasing
temperature. At non-zero temperature the presence of phase-breaking scattering
means that the relevant regime for transport is intermediate between the
coherent and incoherent limits. It is in general very difficult to treat this
regime, but by considering both coherent and incoherent limits we are able to
gain some insight into the behaviour of experimentally relevant cases. 

\subsection{Electronic states in a ballistic domain wall}
\label{sub:ballistic wall states}

For our model of a domain wall we assume a linear rotation of $\vec{f} (\vec{r})
\equiv \vec{f}(z)$ in the $yz$ plane over the region $0 \le z \le \lambda$,
described by
\begin{equation} \label{eq:domain wall profile}
\vec{f}(z) = \left\{
\begin{array}{rl}
(0,\, \sin[\theta(z)],\, -\cos[\theta(z)]), &
0 \le z \le \lambda,\\[1.5mm]
(0,\, 0,\, \sgn[z]), & \textrm{otherwise},
\end{array}
\right.
\end{equation}
where $\theta(z) = \pi z/\lambda$ is the angle between the magnetization inside
the wall and the $z$-axis. Using this profile, one can describe the qualitative
features of transport through a domain wall, which have been shown to be
independent of the detailed form of the wall. \cite{gopar2004} In addition, the
profile of \eqref{eq:domain wall profile} has the advantage that the
corresponding basis states can be found in closed form. \cite{brataas1999,
gopar2004} The local spin eigenstates $\ket{\pm}$ (parallel and anti-parallel to
the direction of $\vec{f}(z)$) are position-dependent inside the domain wall.
Denoting the spin basis states with respect to the fixed $z$-axis as
$\ket{\pm}_z$, we have
\begin{equation}
\ket{\sigma} = R[\theta(z)]\ket{\sigma}_z,
\end{equation}
where 
\begin{equation}
\label{eq:rotation matrix}
R[\theta(z)] 
= \mymatrix{\cos[\theta(z)/2]}{-\sin[\theta(z)/2]}
           {\sin[\theta(z)/2]}{\phantom{-}\cos[\theta(z)/2]}
\end{equation}
is the spinor rotation operator containing the amplitudes
$\braket{\sigma'}{\sigma}_z$ representing the transformation from fixed to local
basis. Because the states $\ket{\pm}$ are position-dependent inside the wall,
the eigenfunctions of \eqref{eq:hamiltonian} in this region are not spin
eigenstates, but a combination of both up and down components:
\begin{equation}\label{eq:spiral basis state}
\tilde{\psi}_{\sigma n}^\gtrless(\vec{r}) = 
\sqrt{\frac{\hbar}{m\tilde{v}_{\sigma n}}}
\rme^{\pm \rmi \tilde{k}_{\sigma n} z}
\phi_n(y) \left( \ket{\sigma} \pm \rmi A_{\sigma n} \ket{-\sigma} \right).
\end{equation}
Here the symbols $\gtrless$ denote the direction of electron propagation (\ie
right or left), while $\sigma=\pm$ is a quantum number representing the spin of
the state $\psi_{\sigma n}(\vec{r})$ to which $\tilde{\psi}_{\sigma n}(\vec{r})$
reduces in the limit $\lambda \rightarrow \infty$. The other parameters are
\begin{subequations}
\begin{eqnarray} \label{eq:dw basis fn parameters}
\tilde{k}_{\sigma n} &=& \sqrt{k_{0 n}^2 + k_\lambda^2 + \frac{\sigma}{2}
\sqrt{k_\Delta^4 + 16 k_\lambda^2 k_{0 n}^2}}, \label{eq:k spiral} \\
A_{\sigma n} &=& \sigma\frac{2k_\lambda\tilde{k}_{\sigma n}}
{\tilde{k}_{\sigma n}^2 + k_\lambda^2 - k_{-\sigma n}^2}, \\
\tilde{v}_{\sigma n} &=& \frac{\hbar \tilde{k}_{\sigma n}}{m}
\left(1+A_{\sigma n}^2 - \sigma 
\frac{2k_\lambda A_{\sigma n}}{\tilde{k}_{\sigma n}}\right),
\end{eqnarray}
\end{subequations}
with $k_{0 n} = \sqrt{(k_{+,n}^2 + k_{-,n}^2)/2}$ and $k_\lambda =
\pi/2\lambda$. Note that the relation between the velocity factor $\tilde{v}
_{\sigma n}$ and wavevector $\tilde{k}_{\sigma n}$ is more complicated than for
the uniform basis states in \eqref{eq:plain basis states}.

The number of propagating modes per spin quantum number inside the domain wall
region, $\tilde{N}_\sigma$, is given by the maximum value of $n$ for which
$\tilde{k}_{\sigma n}$ is real-valued. For parameter values of interest we
generally have $\tilde{N}_\sigma = N_\sigma$. However, for very small $\lambda$
the rotating potential can lead to effective band gaps (analogous to those
occurring in spin-independent periodic potentials), leading to $\tilde{N}
_\sigma < N_\sigma$.
%
%\footnote{A specific example (which does not, however, correspond to the
%parameters used for the calculations in this paper) is given by $\EF=10\eV$,
%$\Delta=1\eV$, $\lambda=1\nanometer$, $L_y=10\nanometer$. For this case
%$N_\pm=53,50$ while $\tilde{N}_\pm=52,50$.} 
%
An additional complication for small $\lambda$ is that states with complex
$\tilde{k}_{\sigma n}$, which are inconvenient to treat within our scattering
matrix approach, may become relevant.
\footnote{ The necessary condition for complex $\tilde{k}_{\sigma n}$ is that
$k_{\sigma n}$ is real (so that the $(\sigma,n)$ channel is open) while $k_{0n}$
is imaginary, such that the term inside the second square root in \eqref{eq:k
spiral} is negative. This can only occur if $k_\Delta < \sqrt{8}k_\lambda$,
which corresponds to a wall width narrower than a single Larmor precession
length and is therefore not applicable for the systems we are considering in
this paper.}
Fortunately, such small values of $\lambda$ do not fall within the range of
parameter values for the systems we are considering, and so in our calculations
we always have $\tilde{N}_\sigma = N_\sigma$ with no complex wavevectors.

In the domain wall model defined by \eqref{eq:domain wall profile}, electrons
incident from the left at $z=0$ or from the right at $z=\lambda$ are scattered
due to the change from uniform to rotating magnetization. The scattering matrices
for these interfaces, which we write $\tilde{s}^{(\mathrm{L})}$ and
$\tilde{s}^{(\mathrm{R})}$, can be calculated using the standard method of matching
incident and scattered wavefunction components. Since the domain wall potential
is uniform in the transverse direction (\ie has no $y$ dependence), the
interface does not mix different transverse modes. We thus have
$\xi^{(\mathrm{L,R})}_{\sigma' n';\sigma n} = 0$ if $n' \ne n$,
for $\xi=t,r,t',r'$. The diagonal amplitudes $\xi ^{(\mathrm{L,R})}
_{\sigma'n;\sigma n}$ are determined by calculating the appropriate scattering
state solutions for states incident on the interface, leading to $4\times4$ sets
of linear equations which we present in Appendix \ref{app:dw S matrix}
(Eqs.~(\ref{eq:left interface matching left}--\ref{eq:left interface matching
right})). 

For general parameter values it is most convenient to solve Eqs.~(\ref{eq:left
interface matching left}--\ref{eq:left interface matching right}) numerically.
However, simple asymptotic expansions can be found in the limits of wide
\cite{waintal2004} and narrow \cite{gopar2004} walls which provide useful
insight. In the experimentally relevant case of small spin-splitting and large
$\lambda$, a particularly simple solution can be found by expanding to second
order in the inverse of the dimensionless parameters
\begin{subequations}
\begin{eqnarray} 
p_n = k_\Delta^2/2k_{0n}k_\lambda, \label{eq:pn def} \\
\pF = k_\Delta^2/2\kF k_\lambda. \label{eq:pF def}
\end{eqnarray}
\end{subequations}
These parameters characterize the effective ``width'' of the domain wall for an
electron in transverse channel $n$ ($p_n$) and for the wire as a whole ($\pF$).
For an individual channel $n$, the adiabatic limit corresponds to $p_n\gg1$ and
is most readily obtained for channels with large $n$ (and hence small $k_{0n}$).
As discussed in Ref.~\onlinecite{gopar2004}, the degree of adiabaticity is in
general channel-dependent. Nevertheless, $\pF$ permits a characterization of
adiabaticity for all states: for a wide wall we have $\pF\gg1$, and hence
$p_n\gg1$ for all channels. Within this assumption we solve Eqs.~(\ref{eq:left
interface matching left}--\ref{eq:left interface matching right}) to
$O(1/p_n^2)$ for each $n$, obtaining
\begin{subequations}\label{eq:interface asymptotic}
\begin{eqnarray} 
\tilde{r}^{(\mathrm{L,R})}_{\sigma' n';\sigma n} &=& 
\tilde{r}'^{(\mathrm{L,R})}_{\sigma' n';\sigma n} = 0, \\
\tilde{t}^{(\mathrm{L,R})}_{\sigma n';\sigma n} 
&=& \tilde{t}'^{(\mathrm{L,R})}_{\sigma n';\sigma n} 
= \delta_{n'n} \left(1-\frac{1}{2p_n^2}\right),\\
\tilde{t}^{(\mathrm{L,R})}_{-\sigma n';\sigma n} 
&=& - \tilde{t}'^{(\mathrm{L,R})}_{-\sigma n';\sigma n} 
= \mp\delta_{n'n}\frac{\rmi}{p_n}.
\end{eqnarray}
\end{subequations}
\eqsref{eq:interface asymptotic} show that there is essentially no reflection
for electrons incident on the domain wall, but that the main effect of the
interfaces is to scatter electrons into a superposition of up and down
transmitted channels (conserving the mode number), with amplitude determined by
$p_n$.

In the regime of small splitting, $\Delta\ll\EF$, the transport properties of
the domain wall are determined primarily by the parameter $\pF$, while the
dependence on $\EF$, $\Delta$, and $\lambda$ individually can, to a good
approximation, be neglected. This is true both for the intrinsic domain wall
scattering shown in \eqsref{eq:interface asymptotic} and for the impurity
scattering to be discussed below. For large splitting, the dependence on
$\Delta$ becomes important as the difference in the number of up and down
conducting channels, $N_+-N_-$, becomes significant. In this work, however, we
are interested only in the case of weak splitting which is relevant for
transition metal ferromagnets. 

\subsubsection{Coherent and incoherent conductance for a ballistic wall}
\label{subsub:ballistic wall}

Before introducing disorder in the following section, we now briefly consider
the case of a disorder-free ballistic domain wall ($V(\vec{r})=0$), which
can be treated by setting $\Nimp=0$ in \eqref{eq:sdw comb coherent} (coherent
case) or \eqref{eq:sdw comb incoherent} (incoherent case). An incoherent
ballistic system represents an idealized scenario in which electrons are subject
to phase-breaking events which do not affect momentum. \cite{datta1997} As such,
it is a useful way to treat the effects of decoherence in mesoscopic transport.
In the case of a domain wall, it is important to understand the differences
between incoherent and coherent transmission in the ballistic case, as this
gives insight into effects which are also relevant for the disordered case.

Within the asymptotic approximation of \eqsref{eq:interface asymptotic}, the
transmission probabilities for coherent and incoherent ballistic walls are, to
$O(1/p_n^2)$, given by
\begin{subequations}\label{eq:asymptotic wall transmission}
\begin{eqnarray}
\label{eq:asymptotic wall transmission co}
T_{-\sigma n';\sigma n} &=& 
\left\{\begin{array}{ll} 
\delta_{n'n}\frac{4}{p_n^2}\sin^2\left(\frac{p_n\pi}{4}\right),
& \textrm{coherent}, \\[1.5mm]
\delta_{n'n}\frac{2}{p_n^2}, & \textrm{incoherent},
\end{array}\right. \\[2mm]
\label{eq:asymptotic wall transmission inc}
T_{\sigma n';\sigma n} &=& 1 - T_{-\sigma n';\sigma n}.
\end{eqnarray}
\end{subequations}
From \eqsref{eq:asymptotic wall transmission} the basic difference between the
coherent and incoherent cases is a suppression of the oscillatory component of
the spin-dependent transmission in the latter case. This occurs because the
oscillations observed in the coherent case arise from the phase interference
between up and down wall basis state components comprising the electron
scattering state, which is suppressed in the incoherent case. This result
implies that the oscillatory torques exerted by a spin-polarized current on a
domain wall predicted in Ref.~\onlinecite{waintal2004} would be suppressed if
the transport were incoherent.

\begin{figure}
\includegraphics[width=0.45\textwidth]{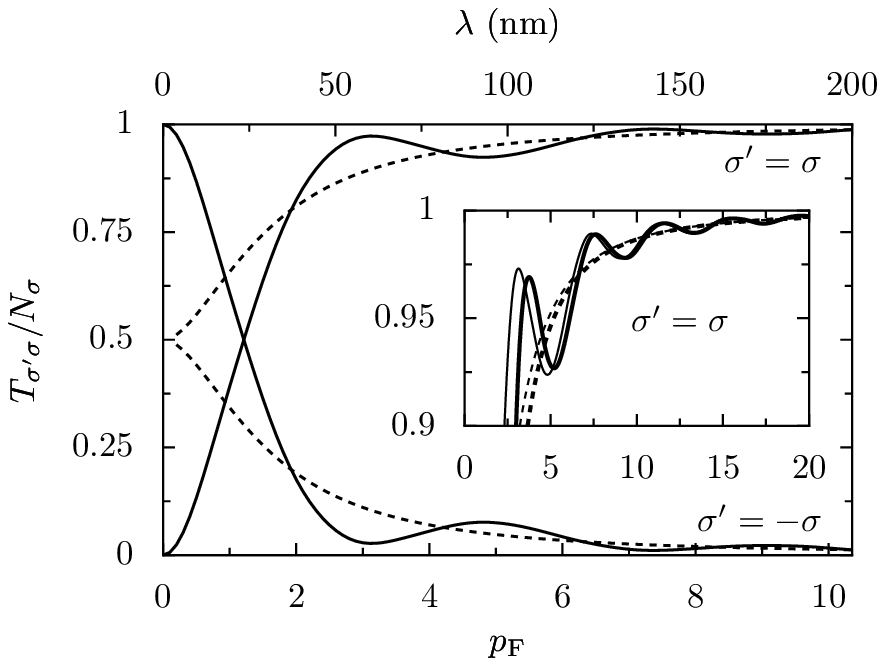}
\caption{\label{fig:ballistic wall}
Spin-dependent transmission per incoming channel, $T_{\sigma'\sigma}/N_\sigma$,
for a ballistic domain wall as a function of the dimensionless effective wall
width $\pF$ (bottom axis). For comparison, the corresponding actual wall width
$\lambda$ for parameter values $\EF=10\eV$ and $\Delta=0.1\eV$ is also shown
(top axis). The curves presented are for $\sigma=+$, with $\sigma'=\pm\sigma$,
and are indistinguishable from the corresponding ones for $\sigma=-$ (for the
parameter values used). Solid and dashed lines indicate, respectively, coherent
and incoherent combination of scattering matrices. The actual wire width used is
$L_y=5\nanometer$, corresponding to $N_\pm=25$, but the curves are
indistinguishable from their form in the limit of large $L_y$. Inset: comparison
of the asymptotic solutions for $T_{\sigma\sigma}$ from \eqref{eq:asymptotic
wall transmission inc} (thick lines) with the exact values (thin lines) over a
larger range of $\pF$.}
\end{figure}

In Figure \ref{fig:ballistic wall} we show the exact total spin-dependent
transmission probabilities $T_{++}$ and $T_{-+}$ (defined in \eqref{eq:total
spin-dep tr}) as a function of $\pF$, for both coherent and incoherent ballistic
walls. The agreement with the asymptotic results from \eqsref{eq:asymptotic wall
transmission} is shown in the inset; it can be seen that these are accurate for
moderate to large $\pF$ values, but for small $\pF$ they diverge significantly.
In the limit $\pF\rightarrow0$, the exact solutions show that the incoherent
transmission goes to $N_\sigma/2$, which is somewhat surprising since we expect
complete mistracking (\ie $T_{\sigma\sigma}=0$) in this limit, as is observed
for the coherent case. The origin of this result can be understood as follows.
By combining scattering matrices incoherently we are effectively ``measuring''
the individual path taken by an electron going through the wall, so that is
projected onto one of the right-moving wall basis states $\psi_\pm^>(\vec{r})$.
Since each of these states comprises an equal weighting of local up and down
components as $\pF\rightarrow0$, the incoherent combination must also yield an
equal weighting. It is clear that the incoherent result is unphysical for small
$\pF$, since there will always be some non-zero distance over which transport is
coherent. We therefore need to be careful when interpreting the incoherent
results for small $\pF$. 

\subsection{Delta function model of scattering}
\label{sub:delta function model}

In transition metal ferromagnets the elastic spin-conserving scattering of
conduction electrons arises from a variety of mechanisms, including impurities,
defects, and grain boundaries. These scattering rates depend on the underlying
details of band structure, and are in general strongly spin-dependent. In spite
of this complex situation, the strength of elastic scattering can be
characterized by only two parameters, namely, the spin-dependent elastic mean
free paths $l_\pm$. The precise microscopic model leading to a given value of
$l_\pm$ is then of minor importance. In the $sd$ model the spin dependence of
$l_\sigma$ arises from two sources: \cite{petej2002} the intrinsic spin
dependence of the electron wavevectors $k_{\sigma n}$ and a difference in the
number of available $d$ states of each spin into which the conducting $s$
electrons can be scattered. The former effect is due to the spin-splitting of
the $s$ band, which makes $k_{+,n} > k_{-,n}$. Therefore, electrons with spin up
are generally less strongly scattered than those with spin down, which in the
absence of other sources of spin dependence leads to $l_+>l_-$. However, since
we work in the regime of small splitting, $\Delta \ll \EF$, this difference is
rather small. The dominant contribution then comes from scattering into the $d$
band. The resulting spin dependence can lead to $l_+ > l_-$ or $l_+ < l_-$
depending on the form of the up and down $d$ sub-bands at the Fermi energy. For
definiteness, we assume that the spin down $d$ sub-band has a greater number of
states at $\EF$ than the spin up sub-band, which implies that spin down
electrons are the more strongly scattered ones, leading to $l_+ > l_-$. To
represent scattering processes, we adopt a simplified picture in which $s$
electrons are scattered by the static potential $V(\vec{r})$. The degrees of
freedom corresponding to the $d$ electrons are therefore not explicitly included
and, in particular, we ignore the possibility that electrons scattered into $d$
states might not return to the $s$ band. 

The potential $V(\vec{r})$ is defined as a random array of delta functions with
spin-dependent amplitudes $u_\pm$, where the orientations are defined with
respect the local magnetization direction $\vec{f}(z)$. The delta function
model is a convenient phenomenological approach to impurity scattering
which has been widely used in spin-independent mesoscopic transport theories
\cite{cahay1988, stone1992} as well as in ferromagnetic systems. \cite{levy1997,
tatara1997} The spin dependence of scattering in the ferromagnet is determined
by the ratio of up and down amplitudes, which we write as
$\rho=u_-/u_+$. For spin-dependent impurity scattering we take $\rho>1$, which
corresponds to $l_+ > l_-$. For completeness, we will also consider the case of
spin-independent impurity scattering, $\rho=1$.

We write the up and down components of $V(\vec{r})$ in the local spin basis as
\begin{equation} \label{eq:Vsigma}
V_\sigma(\vec{r}) = u_\sigma
\sum_{\alpha=1}^{\Nimp}\delta(\vec{r}-\vec{r}_\alpha),
\end{equation}
where there are $\Nimp$ impurities with positions $\vec{r}_\alpha$ randomly
distributed in a region of area $L_y\times L_z$. In general, $L_z$ may be
different from the domain wall length $\lambda$. The impurity density is $\nimp
= \Nimp/L_yL_z$. The total impurity potential can then be written as a sum of
spin-independent and dependent terms: 
\begin{equation} \label{eq:V def1}
V(\vec{r}) = \frac{V_+(\vec{r}) + V_-(\vec{r})}{2} \mathds{1} 
           + \frac{V_+(\vec{r}) - V_-(\vec{r})}{2}
             \vec{f}(z) \cdot \vec{\sigma}.
\end{equation}
Alternatively, the potential can be written in a diagonal form
\begin{equation} \label{eq:V def2}
V(\vec{r}) = 
R^{-1}[\theta(z)]\cdot\mymatrix{V_+(\vec{r})}{0}{0}{V_-(\vec{r})}
\cdot R[\theta(z)],
\end{equation}
in terms of the rotation matrix $R[\theta(z)]$ defined in \eqref{eq:rotation
matrix}.

As shown in Eqs.~(\ref{eq:s comb coherent}--\ref{eq:s comb incoherent}), the
transmission through the disorder potential can be calculated by combining the
scattering matrices $s_\alpha$, $\tilde{s}_\alpha$ of the individual delta
function scatterers located at $\vec{r}_\alpha$, together with the matrices
$p_\alpha$ and $\tilde{p}_\alpha$ for propagation between successive scatterers.
The dependence on the longitudinal component of the scatterer position,
$z_\alpha$, is contained in $p_\alpha$ and $\tilde{p}_\alpha$, while the
dependence on the transverse component, $y_\alpha$, is contained in $s_\alpha$
and $\tilde{s}_\alpha$. We therefore write the scattering matrix for a delta
function scatterer located at $\vec{r}_\alpha$ as $s^{(\delta)}(y_\alpha)$
(uniform) and $\tilde{s}^{(\delta)}(y_\alpha)$ (domain wall). We describe the
calculation of these scattering matrices in Appendix \ref{app:delta function s
matrix} and present solutions valid in the Born approximation in Appendix
\ref{app:Born approximation}.

\begin{figure}
\includegraphics[width=0.45\textwidth]{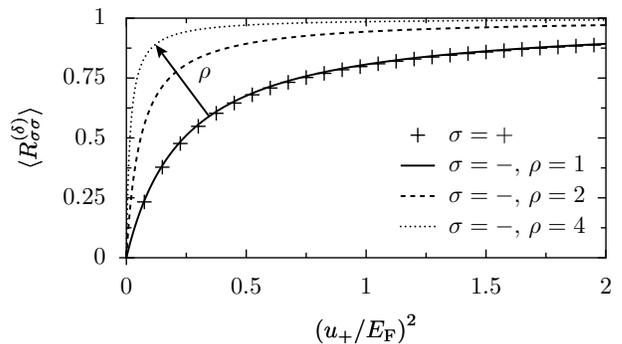}
\caption{\label{fig:delta fn tr}
Total spin-dependent reflection from a delta function scatterer with
spin-dependent amplitudes $u_\sigma$ in the uniformly magnetized case, $\langle
R _{\sigma \sigma} ^{(\delta)} \rangle$ (averaged over transverse positions
$y_\alpha$). Curves are shown as a function of $(u_+/\EF)^2$ for different
values of the up/down scattering ratio $\rho = u_-/u_+$. For $\rho=1$ the only
spin dependence of scattering is due to the small difference between $k_{+,n}$
and $k_{-,n}$, and hence $\langle R_{--}^{(\delta)} \rangle \simeq \langle
R_{++}^{(\delta)} \rangle$. For small $u_\sigma$, $\langle R_{\sigma
\sigma}^{(\delta)} \rangle$ goes as $u_\sigma^2$ (\eqref{eq:born total R
spin-dep uniform}), while for large $u_\sigma$ it approaches a maximum value of
1. The particular system used in this calculation has $L_y=2\nanometer$, giving
$N_\pm=10$.}
\end{figure}

To understand the scattering from a delta function with spin-dependent
amplitudes $u_\sigma$, it is useful to consider the total spin-dependent
reflection for the uniform case, $R_{\sigma \sigma} ^{(\delta)}$. Figure
\ref{fig:delta fn tr} shows $\langle R_{\sigma \sigma} ^{(\delta)} \rangle$, the
reflection averaged over transverse positions $y_\alpha$ of the scatterer, as a
function of $u_\sigma$ for several values of $\rho$. For small $u_\sigma$, we
see that $\langle R_{\sigma \sigma} ^{(\delta)} \rangle$ goes as $u_\sigma^2$,
which can also be seen directly from the Born approximation.
\footnote{For a single delta function, $\langle R_{\sigma \sigma} ^{(\delta)}
\rangle$ can be found from \eqref{eq:born total R spin-dep uniform} after
setting $\Nimp = 1$ and $\nimp = 1/L_yL_z$.}
The ratio of up to down back-scattering thus goes as $\rho^2$ for small $u_\sigma$.
For larger $u_\sigma$, however, $\langle R_{\sigma \sigma}  ^{(\delta)} \rangle$
is bounded above by 1, attaining this value in the limit $u_\sigma \to \infty$.
This occurs because a delta function scatterer can ``block'' at most one
channel.
\footnote{This can be understood intuitively by considering the delta function
as an obstacle in real space. Since the delta function is narrower than a
transverse channel in real-space (whose width is on the order of $1/\kF$), the
maximum reflection corresponds to one channel being completely blocked.}
The spin dependence of reflection is then dramatically reduced for large
$u_\sigma$. Since our goal in this work is a model with spin-dependent
scattering, we must therefore work in the regime of small $u_\sigma$. However, we
cannot take $u_\sigma$ arbitrarily small, since decreasing $u_\sigma$ means that
a larger $N_\sigma$ is required to achieve a given disorder density, leading to
an increase in computation time. In practice we have found $u_+ \lesssim \EF/5$
to be a useful compromise for parameters in the range $1\le \rho \le 2$. 

\section{Transport through a disordered domain wall: intrinsic properties}
\label{sect:intrinsic}

In this section we consider the intrinsic conductance properties of a disordered
domain wall. Here the domain wall constitutes the entire system of interest, \ie
$L_z = \lambda$, and is therefore assumed to be connected on either side to
perfect (disorder-free) leads. By comparing with a uniformly magnetized
disordered region of the same length, we calculate an intrinsic domain wall
magnetoconductance which measures the difference in conductance due to the
presence of a domain wall. We consider both coherent and incoherent regimes; the
corresponding scattering matrices for the system are obtained from
Eqs.~(\ref{eq:s comb coherent}--\ref{eq:s comb incoherent}). Our primary aim in
this section is to study the role of impurity scattering inside the domain wall.
The case of most interest is spin-dependent disorder ($\rho>1$), however, in
order to relate our model to previous works and more clearly understand aspects
of the spin-dependent case, we also consider spin-independent disorder,
$\rho=1$.

In the uniform case, the two spin sub-bands of conduction electrons are
uncoupled, so there is no spin mixing. All off-diagonal (in spin quantum
numbers) transmission and reflection amplitudes are therefore zero, \ie
$\xi_{-\sigma n';\sigma n} = 0$ for $\xi=t,r,t',r'$. The total spin-dependent
transmission and reflection probabilities (defined in \eqsref{eq:summed tr
probabilities}) satisfy
\begin{subequations}
\begin{eqnarray}
\Xi_{-\sigma\sigma}^{(\uni)} = 0, \quad
\Xi_\sigma^{(\uni)} = \Xi_{\sigma\sigma}^{(\uni)},
\end{eqnarray}
\end{subequations}
for $\Xi = T,R,T',R'$.

The conductance properties of the up and down sub-bands of the uniform system
are well described by random matrix theory, \cite{mello1991, beenakker1997}
whose expressions for the average transmission and its moments are in excellent
quantitative agreement with the calculations performed with our model. For a
system of length $L_z$, the total transmission for spin sub-band $\sigma$ in the
incoherent case is given by
\begin{equation} \label{eq:drude}
\langle T_{\sigma,\inc}^{(\uni)} \rangle = N_\sigma/(1+L_z/l_\sigma),
\end{equation}
where $N_\sigma$ is the number of channels and $l_\sigma$ is the spin-dependent
elastic mean free path. \eqref{eq:drude} expresses the transmission as a series
combination of the ballistic contact transmission, $N_\sigma$, with the Drude
conductance, $N_\sigma l_\sigma/L_z$. In the diffusive regime, $l_\sigma \ll
L_z$, the Drude term dominates and $\langle T_{\sigma,\inc}^{(\uni)} \rangle$
displays an Ohmic $1/L_z$ dependence. In the quasi-ballistic regime, $l_\sigma
\gg L_z$, \eqref{eq:drude} is only approximately correct, but is within several
percent of the value obtained by a more precise calculation. \cite{dejong1994a}

For delta function scatterers in a two-dimensional quasi-1D geometry, the mean
free path $l_\sigma$ appearing in \eqref{eq:drude} is given by
\begin{equation}
\label{eq:mfp 2D}
l_\sigma = \frac{2\hbar^3v_{\sigma,\mathrm{F}}}{m\nimp u_\sigma^2},
\end{equation} 
where $v_{\sigma, \mathrm{F}} = \hbar k_{\sigma, \mathrm{F}}/m$ is the Fermi
velocity in spin sub-band $\sigma$. This definition is a factor of 2 larger than
the two-dimensional form \cite{richter1996} since in a quasi-1D geometry forward
scattering processes do not reduce transmission. \cite{baranger1990}

The mean free path $l_\sigma$ can be interpreted intuitively as the typical
distance travelled by an electron of spin $\sigma$ before undergoing a momentum
randomizing scattering event. However, it should be noted that such a conceptual
scattering event does not correspond to the scattering from an individual delta
function scatterer in our model: the length $l_\sigma$ typically corresponds to a
large number of individual scatterers.

The coherent transmission, $\langle T_{\pm, \co}^{(\uni)} \rangle$, is reduced with
respect to  $\langle T_{\pm, \inc}^{(\uni)} \rangle$ by the weak localization
correction, which in a quasi-1D system has the limiting value $-1/3$ in the
diffusive regime $L_z/l_\sigma\gg1$. \cite{mello1991, beenakker1997}

\subsection{Disorder-induced enhancement of spin-mistracking in domain wall
transport coefficients} \label{sub:spin-dep}

\begin{figure}
\includegraphics[width=0.45\textwidth]{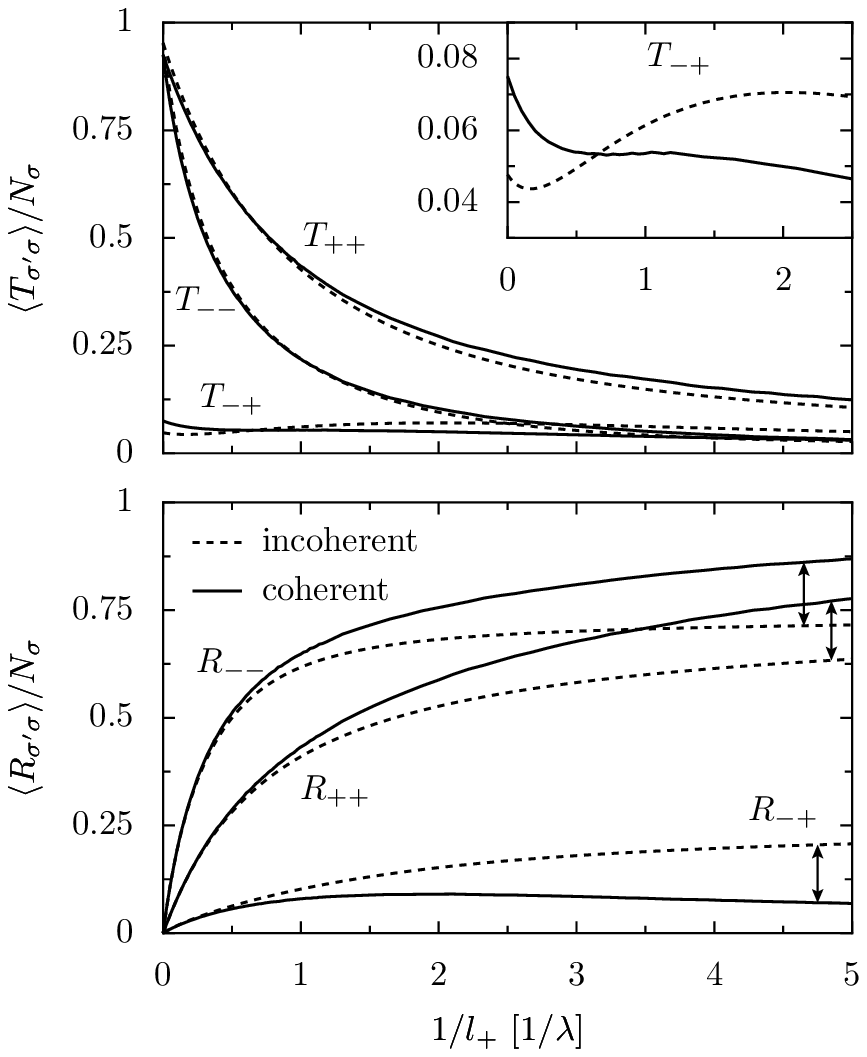}
\caption{\label{fig:dw tr spin-dep}
Spin-dependent domain wall transmission and reflection, $\langle T_{\sigma'
\sigma} \rangle$ and $\langle R_{\sigma' \sigma} \rangle$, (normalized by the
number of conducting modes per incident spin channel, $N_\sigma$) as a function
of disorder, measured by $1/l_+$ (shown in units of $1/\lambda$). Coherent
(solid lines) and incoherent (dashed lines) cases are shown (the arrows connect
corresponding values). The wall width is $\pF = 5$, which corresponds to a wide
wall close to the adiabatic limit ($T_{\sigma \sigma} \gg T_{-\sigma \sigma}$
for the ballistic wall at $1/l_+ = 0$). The parameters for the delta function
scatterers are $u_+=\EF/5$ and $\rho=2$, and the averages are performed using
$N_\mathrm{S} = 2000$ impurity configurations. The wire width is $L_y = 10
\nanometer$, leading to $N_\pm = 51$.}
\end{figure}

We first consider how the transport coefficients of a domain wall vary as a
function of disorder. It is most instructive to study the total spin-dependent
transmission and reflection, summed over all transverse channels and averaged
over impurity configurations, $\langle T_{\sigma' \sigma }^{(\dw)} \rangle$ and
$\langle R_{\sigma' \sigma} ^{(\dw)} \rangle$. In Figure \ref{fig:dw tr spin-dep}
these quantities are shown for a relatively wide wall ($\pF=5$) with
spin-dependent disorder ($\rho = 2$) as a function of disorder strength ($1/l_+$
in units of $1/\lambda$). Both coherent (\eqref{eq:sdw comb coherent}) and
incoherent cases (\eqref{eq:sdw comb incoherent}) are shown. 

At zero disorder, $1/l_+ = 0$, transport through the ballistic wall is highly
adiabatic, so that $\langle T_{\sigma\sigma} ^{(\dw)} \rangle \gg \langle
T_{-\sigma\sigma}^{(\dw)} \rangle$. With increasing disorder, the dominant
transmission coefficient, $\langle T_{\sigma\sigma}^{(\dw)} \rangle$, decreases
quite rapidly, with a form similar to that of the transmission in the uniform
case, $\langle T^{(\uni)}_{\sigma\sigma} \rangle$ (\eqref{eq:drude}). Also, we
observe that $\langle T_{--}^{(\dw)} \rangle$ decreases more rapidly than
$\langle T_{++}^{(\dw)} \rangle$, since $l_-<l_+$. These initial decays can be
quantitatively described by the Born approximation of Appendix \ref{app:Born
approximation}, and the coherent and incoherent cases yield very similar
results. 

The off-diagonal (in spin) coefficients $\langle T^{(\dw)}_{-\sigma\sigma}
\rangle$ also exhibit a linear decrease with $1/l_+$ in the quasi-ballistic
regime. However, for a relatively weak disorder ($\lambda/l_+ \simeq 1$) the
behaviour of off-diagonal coefficients begins to differ considerably from that
of the diagonal ones. As we can see in the inset of the upper graph in Figure
\ref{fig:dw tr spin-dep}, the negative slope of the coherent $\langle T_{-\sigma
\sigma} ^{(\dw)} \rangle$ levels off, and the incoherent $\langle T_{-\sigma
\sigma} ^{(\dw)} \rangle$ increases with disorder. 

In both cases (coherent and incoherent), the magnitude of $\langle
T_{-\sigma\sigma}^{(\dw)} \rangle$ remains relatively constant as a function of
disorder. This means that the \emph{relative} transmission with
spin-mistracking, $\langle T_{-\sigma \sigma} ^{(\dw)} \rangle/ (\langle
T_{\sigma\sigma} ^{(\dw)} \rangle + \langle T_{-\sigma \sigma} ^{(\dw)}
\rangle)$, increases dramatically as a function of disorder. Furthermore, since
$\langle T_{--}^{(\dw)} \rangle < \langle T_{++}^{(\dw)} \rangle$ while $\langle
T_{+-}^{(\dw)} \rangle = \langle T_{-+}^{(\dw)} \rangle$, \emph{the proportion
of transmission with mistracking is greater for the spin down than for the spin
up sub-band}. 

The case of spin-independent disorder, $\rho = 1$, (not shown) is qualitatively
similar to the spin-dependent case just discussed, with the difference that
$\langle T_{++} \rangle \simeq \langle T_{--} \rangle$ and $\langle R_{++}
\rangle \simeq \langle R_{--} \rangle$. This is because in this case the only
spin dependence of the scattering arises from the small spin dependence of the
wavevectors $\tilde{k}_{\sigma n}$.

The preceding observations imply that the adiabaticity which applies to
ballistic domain wall transport no longer applies in the presence of disorder.
In particular, \emph{a wall which is highly adiabatic in the regime of ballistic
transport becomes less so in the presence of disorder}. For spin-independent
disorder, a similar result was recently obtained in the context of transport
through disordered wires in the presence of inhomogeneous magnetic fields.
\cite{popp2003} For spin-dependent disorder, the reduction in adiabaticity
depends on the spin direction, so the wall can no longer be characterized by a
single adiabaticity parameter as it was in the ballistic case.
\cite{falloon2004} The origin of the enhancement of mistracking with increasing
disorder can be understood intuitively as the cumulative effect of small amounts
of mistracking acquired in scattering from each individual delta function (see
Eqs.~(\ref{eq:delta born spiral}b,d)). We illustrate this important idea with a
simple one-dimensional toy model in Appendix \ref{app:spin-mixing 1D}.

Finally, we note several differences between the coherent and incoherent cases
in Figure \ref{fig:dw tr spin-dep}. In the ballistic case, $1/l_+ = 0$, there
is a small difference between $\langle T_{\sigma' \sigma, \co} ^{(\dw)} \rangle$
and $\langle T_{\sigma' \sigma, \inc} ^{(\dw)} \rangle$. This is due to
suppression of the oscillatory component of the coherent transmission, which was
discussed in Section \ref{subsub:ballistic wall} and illustrated in Figure
\ref{fig:ballistic wall}. This difference may be positive or negative, depending
on $\pF$, and becomes increasingly significant as $\pF \rightarrow 0$. The
effect persists for small $1/l_+$, but disappears for larger disorder
since the phase information corresponding to the precessional component is lost
after many scattering events. 

A second difference between the coherent and incoherent cases is that for large
disorder the scattering coefficients with mistracking, $\langle T_{-\sigma
\sigma}^{(\dw)} \rangle$ and $\langle R_{-\sigma \sigma}^{(\dw)} \rangle$, are
smaller for the coherent case than for the incoherent one. The precise origin of
this difference is not clear, but we can eliminate several possible reasons.
Firstly, by looking at the equivalent curves for different values of $\pF$, it
is found that the difference has a constant sign for all $\lambda$. This
suggests that it cannot be explained by the suppression of the precessional
component of transmission in the incoherent case, as invoked in the discussion
of the previous paragraph. Furthermore, it is found that the difference scales
linearly with the number of channels $N_\sigma$ (or equivalently $L_y$). It
cannot, therefore, be explained as a weak localization effect, which should be
characterized by a constant magnitude, independent of $L_y$.

\subsection{Intrinsic domain wall magnetoconductance: sign reversal in coherent
case}
\label{sub:intrinsic dg}

We define the intrinsic domain wall magnetoconductance as the difference between
the conductance of a uniformly magnetized region (``uni'') and that of a domain
wall (``dw'') of equal length with identical impurity configurations 
\begin{equation} \label{eq:dg definition}
\Delta g = g_\uni - g_\dw.
\end{equation}
Assuming that we can contact the wall directly, the impurity average $\langle
\Delta g \rangle$ could be associated with the change in the measured
conductance when an external magnetic field along the $z$ direction is applied
in such a way as to destroy the domain wall and arrive at a magnetically
homogeneous configuration. Although they are not of direct experimental
relevance, we also define the differences in spin-dependent transmission,
$\Delta T_\sigma = T_\sigma ^{(\uni)} - T_\sigma^{(\dw)}$ in order to guide our
physical discussion. With the above notations we obviously have $\Delta g =
\Delta T_+ +\Delta T_-$.

\begin{figure}
\includegraphics[width=0.45\textwidth]{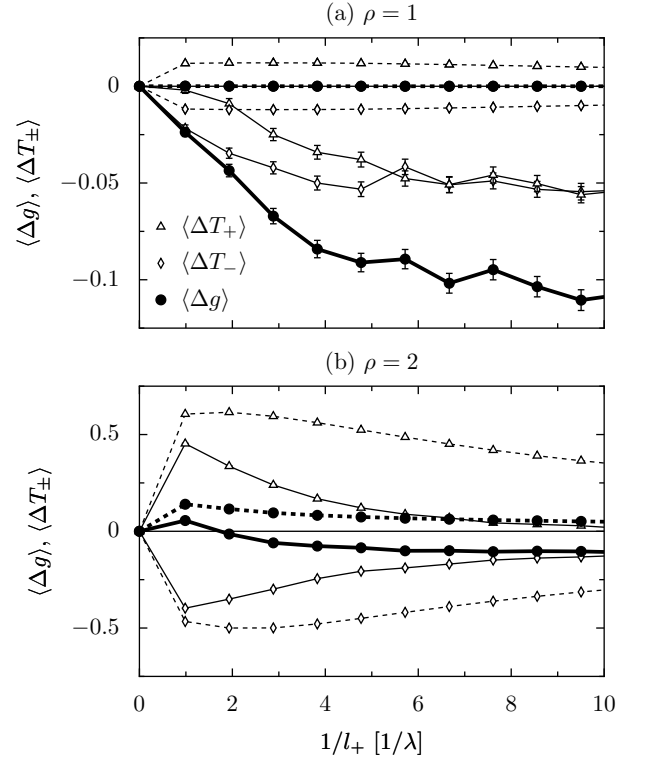}
\caption{\label{fig:delta G vs disorder}
Coherent (solid lines) and incoherent (dashed lines) domain wall
magnetoconductance $\langle \Delta g \rangle$ (filled symbols) as a function of
$1/l_+$ (in units of $1/\lambda$) for a domain wall with $\pF=5$, in the case of
(a) spin-independent disorder ($\rho=1$) and (b) spin-dependent disorder
($\rho=2$). The differences in spin-dependent transmission $\langle T_\pm
\rangle$ are also shown (empty symbols). The system parameters are as for Figure
\ref{fig:dw tr spin-dep}, with $N_\mathrm{S}=3000$ impurity configurations.}
\end{figure}

In Figure \ref{fig:delta G vs disorder} we show the disorder-averaged $\langle
\Delta g \rangle$ (thick lines, filled symbols) and $\langle \Delta T_\pm
\rangle$ (thin lines, empty symbols) as a function of disorder, for
spin-independent ($\rho=1$) and spin-dependent ($\rho=2$) disorder, in both
coherent and incoherent cases. 

In the spin-independent case ($\rho=1$, Figure \ref{fig:delta G vs disorder}a),
the main feature is \emph{a negative coherent magnetoconductance, which becomes
positive and very small in the incoherent regime}. For the spin-dependent case
($\rho=2$, Figure \ref{fig:delta G vs disorder}b) \emph{the negative coherent
magnetoconductance is obtained above a threshold disorder, and a positive
$\langle \Delta g_\co \rangle$ appears in the incoherent regime}. Below we
comment on the generality of these basic findings, their relationship with
previously found effects, and their physical relevance.

A negative coherent magnetoconductance was predicted by Tatara and Fukuyama
\cite{tatara1997, tatara2001} for the case of spin-independent disorder, and
interpreted as a weak localization effect. Such an effect has, however, eluded
experimental confirmation. The underlying reason for the putative reduction of
weak localization in a domain wall is a suppression of backscattering processes
which conserve spin. That is, $r_{\sigma n';\sigma n}$ is reduced due to the
possibility of scattering into the opposite spin channel, $r_{-\sigma n';\sigma
n}$. Since weak localization stems from an enhancement of the diagonal
spin-conserving reflection amplitudes $r_{\sigma n;\sigma n}$ (coherent
backscattering), it follows that the effect will be reduced in a domain wall as
compared to a uniformly magnetized region.

Our numerical results suggest that a suppression of weak localization by the
domain wall is indeed the dominant mechanism responsible for the coherent
magnetoconductance in the regime of large disorder. Firstly, the limiting value
$\langle \Delta g_\co \rangle \simeq -0.1$, obtained in the diffusive
regime, is of the order of the quasi-1D weak localization value of $-1/3$ for the
uniform case. \cite{mello1991, beenakker1997} Furthermore, this limiting value is
approximately independent of system size ($L_y$ and $\lambda$),
\footnote{ In the limit $\lambda \rightarrow \infty$, the domain wall becomes
perfectly adiabatic and $\langle \Delta g_\co \rangle$ should go to zero.
However, for the system sizes we are able to treat numerically, this limit
corresponds to values of $\lambda$ much larger than the localization length
($N_\sigma l_\sigma$), for which the system is in the strongly localized regime
and is therefore not relevant for diffusive transport.}
which is a general characteristic of coherence effects such as weak localization.
 
\begin{figure}
\includegraphics[width=0.45\textwidth]{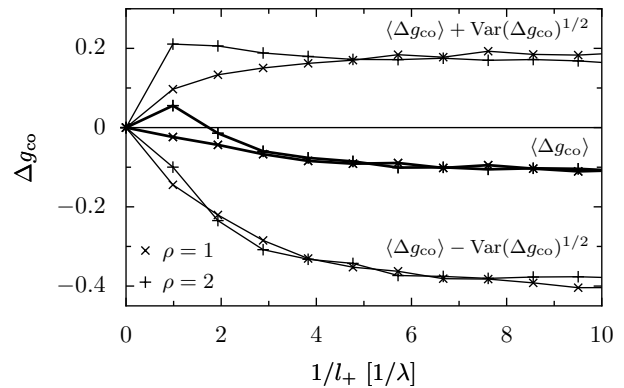}
\caption{\label{fig:delta G distribution}
Distribution of the coherent intrinsic magnetoconductance, $\Delta g_\co$, for
spin-independent ($\rho=1$, $\times$) and spin-dependent ($\rho=2$, $+$)
disorder. The curves in the centre show $\langle \Delta g_\co \rangle$, while
the upper and lower curves show $\langle \Delta g_\co \rangle \pm \var(\Delta
g_\co) ^{1/2}$ respectively. All parameters are as for Figure \ref{fig:delta G
vs disorder}.}
\end{figure}

From the point of view of physically measurable effects, it is important to note
that the coherent magnetoconductance $\langle \Delta g_\co \rangle$ is
characterized by relatively large fluctuations. Indeed, in the diffusive regime
$\Delta g_\co$ follows approximately a normal distribution with fluctuations
characterized by $\sqrt{\var(\Delta g_\co)} = (\langle \Delta g_\co^2 \rangle -
\langle \Delta g_\co \rangle^2)^{1/2} \simeq 0.3$, as illustrated in Figure
\ref{fig:delta G distribution}. For large disorder, the magnitude of these
fluctuations is independent of system size ($L_y$) and up/down scattering ratio
($\rho$), and is related to the universal fluctuations of $g_\dw$ and $g_\uni$.
We discuss these aspects in more detail in the following section. We notice that
$\sqrt{\var(\Delta g_\co)} \simeq 3 \lvert \langle \Delta g_\co \rangle \rvert$
for large disorder. For an individual disorder configuration, there is thus a
significant probability for $\Delta g _\co$ to be positive as well as negative.
This could make it difficult for a negative $\langle \Delta g _\co \rangle$ to
be detected experimentally (even in the coherent regime) beyond statistical
uncertainty.

Another significant factor concerning the coherent magnetoconductance is that in
the case of spin-dependent disorder $\langle \Delta g_\co \rangle$ is positive
for small disorder. The value of $1/l_+$ at which it changes sign increases with
the system width $L_y$. In our calculations, numerical constraints limit us to
systems containing on the order of $10^2$ channels. However, the nanowires of
experiments such as Ref.~\onlinecite{ebels2000} contain on the order of $10^4$
channels. The region in which $\langle \Delta g_\co \rangle$ is negative would
then correspond to an unrealistically large disorder, particularly for the
cobalt nanowires of Ref.~\onlinecite{ebels2000}, in which domain walls are
relatively narrow ($\lambda \simeq 15 \nanometer$). For materials with wider
walls, such as Nickel ($\lambda \simeq 100 \nanometer$), the necessary density
of impurity scatterers would be smaller, although it would be more difficult to
have phase coherence across the greater length of the wall. Thus, to
access the regime of negative $\langle \Delta g_\co \rangle$ experimentally it
would be necessary to work with very small, highly disordered nanowires at low
temperature. 

%It is interesting that for spin-independent disorder (Figure \ref{fig:delta G vs
%disorder}), both $\langle \Delta T_{+,\co} \rangle$ and $\langle \Delta
%T_{-,\co} \rangle$ are negative, with a limiting value of $\sim$$-0.05$ for
%$1/l_+$ large. These combine to give the above-mentioned limiting value $\langle
%\Delta g_\co \rangle \simeq -0.1$. For spin-dependent disorder, a similar
%limiting value of $\langle \Delta g_\co \rangle$ is obtained, but the two values
%$\langle \Delta T_{\pm,\co} \rangle$ have different sign (see Figure
%\ref{fig:delta G vs disorder}b). 
%

In the incoherent regime, the magnetoconductance $\langle \Delta g _\inc
\rangle$ in the case of spin-independent disorder is positive but extremely
small (approximately two orders of magnitude smaller than $\lvert \langle \Delta
T_{\pm,\inc} \rangle \rvert$). For spin-dependent disorder $\langle \Delta g
_\inc \rangle$ is much larger. Although not shown in Figure \ref{fig:delta G vs
disorder}, we note that the incoherent quantities scale linearly with increasing
$L_y$, while for fixed $1/l_+$ they decrease with increasing $\lambda$ (see
Figure \ref{fig:dgong_int vs lambda}). 

The behaviour of $\langle \Delta g_\inc \rangle$ can be understood as a
combination of spin-mistracking with spin-dependent scattering. As we saw in
Section \ref{sub:spin-dep}, successive scattering events in the domain wall
enhance spin-mistracking. This means that \emph{the electron is sensitive not
only to the impurity potential associated with its incoming spin, but also to
the one with the opposite orientation}. Since the scattering strengths are
spin-dependent, this leads either to a reduction or enhancement of transmission
compared to the uniform case, according to the spin direction: $\langle \Delta
T_{+,\inc} \rangle > 0$ and $\langle \Delta T_{-,\inc} \rangle < 0$. For
spin-independent disorder, these differences are small and almost equal in
magnitude, so that $\langle \Delta g_\inc \rangle$ is negligibly small. For
spin-dependent disorder, however, we have $\lvert \langle \Delta T_{+,\inc}
\rangle \rvert > \lvert \langle \Delta T_{-,\inc} \rangle \rvert$, which leads
to a positive and much larger value of $\langle \Delta g_\inc \rangle$. In this
case there is thus a significant magnetoconductance arising from \emph{an
enhancement of back-scattering due to the exposure of electrons in a
superposition of spin states to spin-dependent scattering}. This idea was used
at a phenomenological level in Ref.~\onlinecite{viret1996}, where transport
through a diffusive domain wall in the almost adiabatic limit of small
spin-mistracking was described by a reduced effective mean free path
representing a weighted average between spin up and down mean free paths. 

In the ballistic limit, $1/l_+ \to 0$, there is no impurity scattering inside
the wall. The back-scattering is then entirely due to reflection from the wall
interfaces, which is extremely weak. Both $\langle \Delta g_\co \rangle$ and
$\langle \Delta g_\inc \rangle$ therefore become negligible in this limit. 

In the diffusive regime ($\lambda/l_+ \ll 1$), we believe that two primary
mechanisms explain the different behaviour of $\langle \Delta g_\co \rangle$ and
$\langle \Delta g_\inc \rangle$. Firstly, the limiting value of $\langle \Delta
g _\co \rangle$ for large disorder appears to be due to a reduction of weak
localization by the domain wall, which does not apply in the incoherent case.
Secondly, as we discussed in Section \ref{sub:spin-dep}, the spin-mistracking
transmission and reflection, $\langle T_{-\sigma \sigma, \co}^{(\dw)} \rangle$
and $\langle R_{-\sigma \sigma, \co}^{(\dw)} \rangle$, are reduced in the
coherent regime with respect to the incoherent one. We propose that this leads
to a suppression of the enhanced backscattering in the domain wall which is
dominant in the incoherent case. For large disorder, the positive
magnetoconductance present in the incoherent case is then cancelled out in the
coherent case, leaving only the negative component from the weak localization
reduction.

\subsection{Reduction of universal conductance fluctuations for coherent domain
wall}

An important feature of quantum transport through diffusive coherent systems is
that the conductance fluctuations, $(\delta g^2)^{1/2} = \var(g)^{1/2}$, are
\emph{universal}, with a magnitude of order $e^2/h$ independent of the system
size or mean free path. \cite{lee1985a} For a spinless quasi-1D disordered
system, the conductance fluctuations have the value $\sqrt{2/15}$ in units of
$e^2/h$. In our model this applies to each spin sub-band in the uniformly
magnetized case, so that
\begin{equation} \label{eq:tpm uni ucf}
(\delta[T_\pm^{(\uni)}]^2)^{1/2} = \sqrt{2/15}\,.
\end{equation}

For a non-magnetic system (``nm''), there is no spin-splitting of the $s$ band
(which would correspond to $\Delta = 0$ in our model) and the transport of up
and down electrons is identical. In that case we have $T_+^{(\mathrm{nm})} =
T_-^{(\mathrm{nm})}$ and hence $(\delta[T_+^{(\mathrm{nm})} +
T_-^{(\mathrm{nm})}]^2)^{1/2} = 2 (\delta[T_+^{(\mathrm{nm})}] ^2) ^{1/2}$. The
conductance fluctuations are then
\begin{equation}
(\delta g_\mathrm{nm}^2)^{1/2} 
= 2(\delta[T_+^{(\mathrm{nm})}]^2)^{1/2} 
= \sqrt{8/15}\,.
\end{equation}

In a ferromagnetic system ($\Delta\neq0$), on the other hand, the spin
dependence of the wavevectors $k_{\sigma n}$ suppresses the correlation between
$T_+^{(\uni)}$ and $T_-^{(\uni)}$. In this case we have $(\delta[T_+^{(\uni)} +
T_-^{(\uni)}]^2)^{1/2} = (2\delta [T_+^{(\uni)}]^2) ^{1/2}$, so the conductance
fluctuations are reduced as compared to those of a non-magnetic system:
\begin{equation} \label{eq:g uni ucf}
(\delta g_\uni^2)^{1/2} 
= (2\delta[T_+^{(\uni)}]^2)^{1/2} 
= \sqrt{4/15}\,.
\end{equation}
This reduction due to the non-degeneracy of spin states is directly analogous to
the experimentally observed reduction of conductance fluctuations due to Zeeman
splitting in an applied magnetic field. \cite{beenakker1997} 

\begin{figure}
\includegraphics[width=0.45\textwidth]{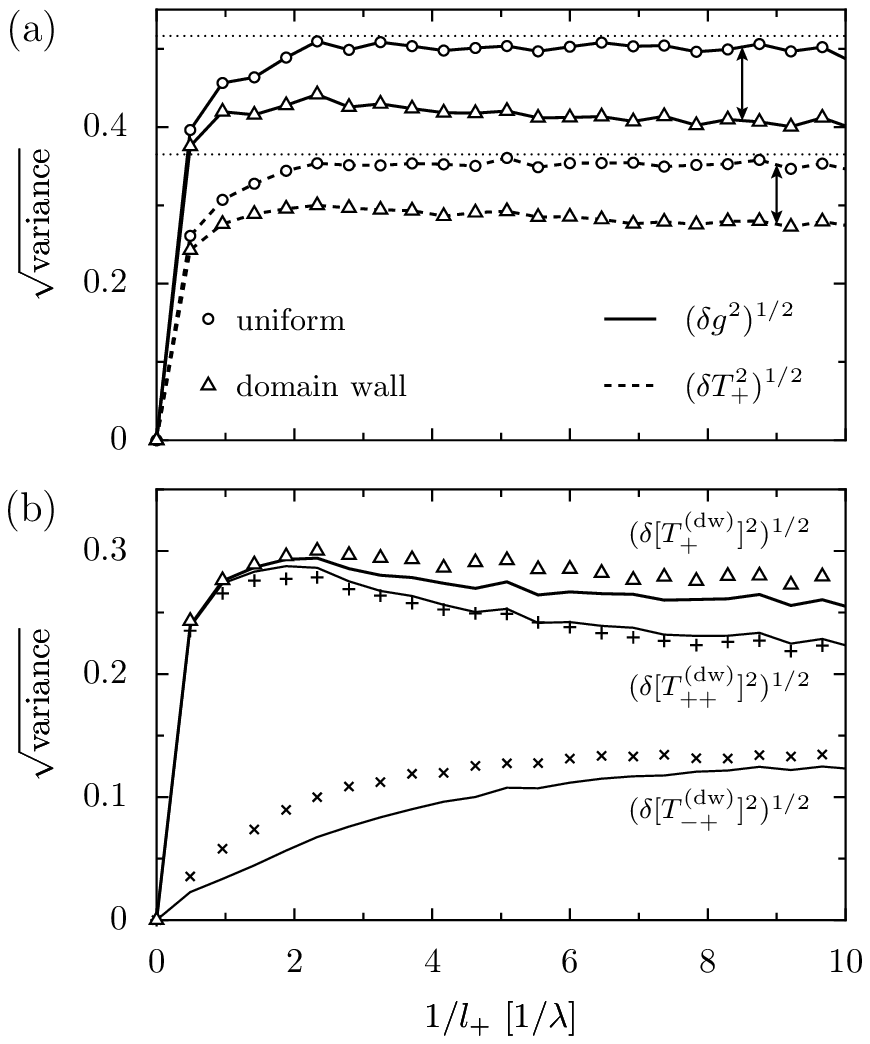}
\caption{\label{fig:ucf}
(a) Fluctuations of conductance, $(\delta g^2)^{1/2}$, (solid lines) and
spin-dependent transmission, $(\delta T_+^2)^{1/2}$, (dashed lines) for a
uniformly magnetized region ($\circ$) and domain wall ($\triangle$), in the case
of spin-independent disorder ($\rho=1$). The fluctuations of spin-down
transmission, $(\delta T_-^2)^{1/2}$ are not shown, but agree with $(\delta
T_+^2)^{1/2}$ to within the statistical error of the calculation. The dotted
horizontal lines indicate the theoretical values in the uniform case, $(\delta
g_\uni^2)^{1/2} = \sqrt{4/15}$ and $(\delta[T_\pm^{(\uni)}]^2) ^{1/2} =
\sqrt{2/15}$, which are in good agreement with our numerical results for
$\lambda/l_+ \gtrsim 2$. (b) Fluctuations of the spin-dependent domain wall
transmission coefficients: $(\delta[T_+^{(\dw)}]^2)^{1/2}$ ($\triangle$),
$(\delta[T_{++}^{(\dw)}]^2) ^{1/2}$ ($+$), and $(\delta[T_{-+}^{(\dw)}]^2)
^{1/2}$ ($\times$). The corresponding estimated values, based on the
approximation in Eqs.~(\ref{eq:dw ucf components}--\ref{eq:dw ucf
decorrelation}), are also shown (solid lines). System parameters are $\pF=5$ and
$L_y=10\nanometer$, with $3000$ impurity configurations.}
\end{figure}

In Figure \ref{fig:ucf}a we show the conductance fluctuations obtained from our
model in the uniform case, $(\delta g_\uni^2)^{1/2}$ and $(\delta[T_\pm
^{(\uni)}] ^2) ^{1/2}$, as a function of $1/l_+$. They are in excellent
agreement with the theoretical values from \eqsref{eq:tpm uni ucf} and
(\ref{eq:g uni ucf}) for $\lambda/l_+ \gtrsim 2$. 

Figure \ref{fig:ucf}a also shows the conductance fluctuations for a domain wall
region, $(\delta g_\dw^2)^{1/2}$, together with $(\delta[T_\pm ^{(\dw)}]^2)
^{1/2}$. We see that \emph{the conductance fluctuations in a disordered domain
wall are reduced with respect to those of a uniformly magnetized region}.
Moreover, the conductance fluctuations are no longer universal since a slow
decrease is observed as a function of $1/l_+$ in the diffusive regime. 

This result can be understood as arising from statistical decorrelation between
the components of $T_\sigma^{(\dw)}$. For the domain wall coefficients
$T_{\sigma'\sigma}^{(\dw)}$ we do not have a way of estimating the fluctuations
analogous to \eqref{eq:tpm uni ucf}. However, we can make the hypothesis that
the relative fluctuations for each component are the same as in the uniform
case:
\begin{equation} \label{eq:dw ucf components}
\frac{(\delta[T_{\sigma'\sigma}^{(\dw)}]^2)^{1/2}}
     {\langle T_{\sigma'\sigma} ^{(\dw)} \rangle} 
\simeq
\frac{(\delta[T_\sigma^{(\uni)}]^2)^{1/2}}
     {\langle T_\sigma ^{(\uni)} \rangle}\,.
\end{equation}
Furthermore, we can expect that $T_{\sigma \sigma} ^{(\dw)}$ and $T_{-\sigma
\sigma} ^{(\dw)}$ will be uncorrelated in the diffusive regime, so that
\begin{equation} \label{eq:dw ucf components uncorrelated}
(\delta[T_\sigma^\mathrm{(dw)}]^2)^{1/2} \simeq (\delta[T _{\sigma\sigma}
^\mathrm{(dw)} ]^2 + \delta[T _{-\sigma\sigma} ^\mathrm{(dw)} ]^2) ^{1/2}.
\end{equation}
Using the fact that $\langle T_\sigma ^\mathrm{(dw)} \rangle = \langle
T_{\sigma\sigma} ^\mathrm{(dw)} \rangle + \langle T_{-\sigma\sigma}
^\mathrm{(dw)} \rangle$, Eqs.~(\ref{eq:dw ucf components}--\ref{eq:dw ucf
components uncorrelated}) then give
\begin{widetext}
\begin{equation} \label{eq:dw ucf decorrelation}
(\delta[T_\sigma^{(\dw)}]^2)^{1/2} 
\simeq 
\frac{\langle T_\sigma^{(\dw)} \rangle}
     {\langle T_\sigma^{(\uni)} \rangle}
\left(1-
\frac{2\langle T_{\sigma\sigma}^{(\dw)} \rangle 
       \langle T_{-\sigma\sigma}^{(\dw)} \rangle}
     {\langle T_\sigma^{(\dw)} \rangle^2}
\right)^{1/2}
(\delta[T_\sigma^{(\uni)}]^2)^{1/2}\,.
\end{equation}
\end{widetext}
The first factor on the r.h.s.\ is approximately unity since $\langle
T_\sigma^{(\dw)} \rangle \simeq \langle T_\sigma^{(\uni)} \rangle$. We thus see
that $(\delta[T_ \sigma ^{(\dw)}] ^2) ^{1/2}$ is reduced with respect to
$(\delta[T_ \sigma ^{(\uni)}] ^2) ^{1/2}$ by the second term on the r.h.s.\ of
\eqref{eq:dw ucf decorrelation}. This reduction factor is most important when
the two transmission coefficients $\langle T_{\sigma\sigma} ^{(\dw)} \rangle$
and $\langle T_{-\sigma\sigma} ^{(\dw)} \rangle$ are of the same order, which is
the situation approached with increasing disorder.

In Figure \ref{fig:ucf}b we compare the actual fluctuations $(\delta[T _\sigma
^{(\dw)}] ^2) ^{1/2}$ and $(\delta[T _{\sigma'\sigma} ^{(\dw)}] ^2) ^{1/2}$ with
the corresponding predictions of \eqref{eq:dw ucf decorrelation}. The two agree
reasonably well and, significantly, the approximate values reproduce the
decreasing behaviour of $(\delta[T _\sigma ^{(\dw)}] ^2) ^{1/2}$ for $1/l_+$
large. The simple hypotheses contained in Eqs.~(\ref{eq:dw ucf
components}--\ref{eq:dw ucf components uncorrelated}) thus explain
qualititatively the two essential features of $(\delta[T _\sigma ^{(\dw)}
]^2)^{1/2}$, which are an overall reduction with respect to $(\delta[T _\sigma
^{(\uni)} ]^2)^{1/2}$ and a loss of universality represented by a slow decrease
with increasing disorder. 

%The discrepancy between exact and estimated fluctuations is largest in
%the case of $T_{-\sigma \sigma} ^{(\dw)}$, for which the actual values are
%larger than the estimated ones. This suggests there is a non-zero correlation
%between $T_{\sigma \sigma} ^{(\dw)}$ and $T_{-\sigma \sigma} ^{(\dw)}$, which
%has a proportionately larger effect on $(\delta[T _{-\sigma \sigma} ^{(\dw)}]^2
%)^{1/2}$. 

%Finally, we comment on the dependence of the domain wall conductance
%fluctuations on the parameters $\lambda$ and $\rho$. Within the diffusive
%regime, it is found that $(\delta g_\dw^2)^{1/2}$ does not vary significantly
%with either $\lambda$ or $\rho$. However, the fluctuations of the spin-dependent
%transmission, $(\delta[T_\pm^{(\dw)}]^2)^{1/2}$, are sensitive to $\rho$. For
%$\rho = 1$ (the case shown in Figure \ref{fig:ucf}), we have $(\delta[T_+
%^{(\dw)}] ^2) ^{1/2} = (\delta[T_- ^{(\dw)}] ^2) ^{1/2}$, while $(\delta[T_+
%^{(\dw)}] ^2) ^{1/2} > (\delta[T_- ^{(\dw)}] ^2) ^{1/2}$ for $\rho > 1$. This
%can be understood once again in terms of the simple interpretation described
%above, since for $\rho > 1$ we have $\langle T_{-+} ^{(\dw)} \rangle / \langle
%T_{++} ^{(\dw)} \rangle > \langle T_{+-} ^{(\dw)} \rangle / \langle T_{--}
%^{(\dw)} \rangle$.

\subsection{Non-monotonic $\lambda$-dependence of incoherent relative
magnetoconductance}
\label{sub:intrinsic dgong}

We now consider the relative intrinsic domain wall magnetoconductance, defined
as
\begin{equation} \label{eq:dgongdw}
\dgongdw = 
\frac{g_\uni - g_\dw}{g_\uni}.
\end{equation}
For disordered systems this is a more useful quantity than the
magnetoconductance $\Delta g$ discussed in Section \ref{sub:intrinsic dg}, as it
measures the \emph{relative} importance of the magnetoconductance effect inside
the wall. In this section we consider only the incoherent regime; for this case,
$\dgongdw$ is independent of $L_y$ since $\langle \Delta g_\inc \rangle$ scales
linearly with transverse system size. 

\begin{figure}
\includegraphics[width=0.45\textwidth]{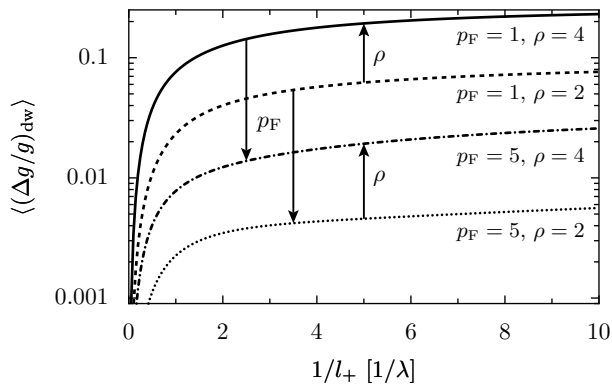}
\caption{\label{fig:dgong_int vs lp}
Intrinsic relative domain wall magnetoconductance, $\langle \dgongdw \rangle$,
as a function of disorder strength, $1/l_+$ (in units of $1/\lambda$), in the
incoherent case. The different curves illustrate the dependence on wall width
($\pF=1,5$) and spin dependence of scattering ($\rho=2,4$). The arrows indicate
the evolution of the curves upon increasing these parameters. Other system
parameters are as for Figure \ref{fig:dw tr spin-dep}.}
\end{figure}

In Figure \ref{fig:dgong_int vs lp} we plot $\langle \dgongdw \rangle$ as a
function of $1/l_+$ (in units of $1/\lambda$). Various domain wall widths ($\pF
= 1, 5$) and up/down scattering ratios ($\rho = 2, 4$) are shown. We see that
\emph{$\langle \dgongdw \rangle$ is a monotonically increasing function of
disorder}. This shows that the decreasing behaviour of $\langle \Delta g_\inc
\rangle$ for large disorder is due simply to the overall decrease in conductance
with disorder. The monotonicity of $\langle \dgongdw \rangle$ occurs because the
amount of spin-mistracking, and hence also the enhancement of back-scattering,
is proportional to the amount of impurity scattering, as discussed in Section
\ref{sub:intrinsic dg}. It can also be seen from Figure \ref{fig:dgong_int vs
lp} that $\langle \dgongdw \rangle$ increases with $\rho$ and decreases with
$\pF$, which is consistent with the explanation of the intrinsic
magnetoconductance given in Section \ref{sub:intrinsic dg}. 

We note that the models of Refs.\ \onlinecite{viret1996} and
\onlinecite{levy1997} assume diffusive transport, so that the results obtained
therein correspond to the regime of large $1/l_+$ in Figure \ref{fig:dgong_int
vs lp}. The magnetoconductance predicted in these works depends on the up/down
scattering ratio $\rho$, but is independent of $l_+$ (provided it satisfies
$l_+\ll\lambda$). This is consistent with the levelling off of the curves in
Figure \ref{fig:dgong_int vs lp} with increasing $1/l_+$, and suggests that
$\langle \dgongdw \rangle$ approaches a finite limit for large disorder.

\begin{figure}
\includegraphics[width=0.45\textwidth]{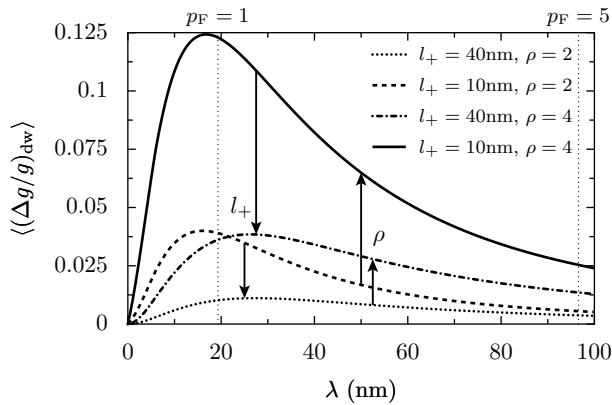}
\caption{\label{fig:dgong_int vs lambda}
Intrinsic relative domain wall magnetoconductance, $\langle \dgongdw \rangle$,
as a function of wall width, $\lambda$, in the incoherent case. Several values
of mean free path ($l_+=10\nanometer, 40\nanometer$) and spin dependence of
scattering ($\rho=2,4$) are shown. The arrows show direction in which the curves
evolve upon increasing these parameters. The two values of wall width
corresponding to Figure \ref{fig:dgong_int vs lp} ($\pF=1,5$) are indicated by
dotted vertical lines. Other system parameters are as for Figure \ref{fig:dw tr
spin-dep}.}
\end{figure}

Figure \ref{fig:dgong_int vs lambda} shows $\langle \dgongdw \rangle$ as a
function of wall width $\lambda$, for $\rho = 2,4$ and $l_+=10\nanometer,
40\nanometer$. For large $\lambda$, we find a $1/\lambda$ dependence in
agreement with Refs.\ \onlinecite{viret1996} and \onlinecite{levy1997}. This can
be understood because with increasing wall width the amount of spin-mistracking
is reduced. In the limit $\lambda \to 0$, on the other hand, the domain wall
becomes an abrupt interface with no impurities and $\langle \dgongdw \rangle$
becomes negligible.
\footnote{When $N_+>N_-$ there is a small interface resistance due to reflection
of spin up electrons with longitudinal energy less than $\Delta$, which gives
rise to a relative magnetoconductance $\Delta g/g \sim \Delta/\EF$.
\cite{weinmann2001, gopar2004} In our calculations this effect is negligible
since we have $N_+ = N_-$.}
In between these two extreme cases, \emph{there must therefore be a maximum
value of $\langle \dgongdw \rangle$ at a finite value of $\lambda$.} This value
represents an optimum intermediate situation: a wall which is wide enough to
contain an appreciable amount of impurity scattering, but narrow enough to cause
significant spin-mistracking. From Figure \ref{fig:dgong_int vs lambda} it can
be inferred that this value of $\lambda$ decreases with increasing disorder (\ie
decreasing $l_\sigma$). On the other hand, the different values of $\rho$ affect
the overall magnitude of $\langle \dgongdw \rangle$, but do not significantly
change the position of its maximum value. 

\section{Domain wall in a nanowire: effect of scattering in the leads}
\label{sect:circuit}

To understand the effect of a domain wall on electron transport in a nanowire,
it is necessary to consider not only the \emph{intrinsic} effects described in
the previous section, but also \emph{extrinsic} effects arising from scattering
in the regions adjacent to the wall. In a previous work \cite{falloon2004} we
analysed such effects in the special case of a ballistic domain wall, using a
circuit model which incorporates spin-dependent scattering in the regions on
either side of the wall using a generalization of the two-resistor model of GMR.
\cite{valet1993} The essence of this model is to treat transport of electrons
with different spins in the adjacent regions as independent over a distance
equal to the spin-diffusion length, $\lsd$. This is the length scale over which
the distribution between the two spin sub-bands is equilibriated through
spin-flip scattering processes. In Ref.~\onlinecite{falloon2004}, transport in
the adjacent regions was treated using classical spin-dependent resistors. Such
an approach is valid under the assumption that the phase coherence length,
$L_\phi$, satisfies $\lambda\lesssim L_\phi \ll \lsd$. 

\begin{figure}
\includegraphics[width=0.45\textwidth]{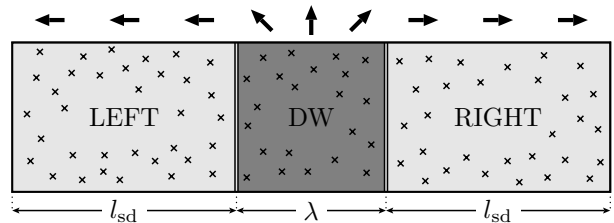}
\caption{\label{fig:circuit model}
Diagram illustrating the physical region considered in Section
\ref{sect:circuit}. On either side of the domain wall are uniformly magnetized
regions of length $\lsd$ representing the distance over which electrons
propagate before spin relaxation.}
\end{figure}

In this section we follow a similar method to incorporate extrinsic scattering
(outside the domain wall region) into our model. However, in place of classical
spin-dependent resistors, we treat the impurity scattering in the regions
adjacent to the domain wall using the same delta function scatterer model as
inside the domain wall. The resulting model system is illustrated in Figure
\ref{fig:circuit model}. This approach differs with respect to
Ref.~\onlinecite{falloon2004} in that the adjacent regions are combined with the
domain wall using the scattering matrix combination formula (\eqsref{eq:S matrix
combination}), rather than combining spin-dependent conductances using
Kirchhoff's rules. We are interested in the limit $l_\pm \ll \lsd$, for which
the difference between these two approaches is not significant. However, the
present approach has the advantage that it remains valid in principle for any
value of $l_\pm/\lsd$. In addition, phase coherence lengths $L_\phi > \lambda$
can be considered within this approach (although we do not do so in this paper). 

As in Section \ref{sect:intrinsic}, we consider conductance through the domain
wall in both coherent and incoherent cases. In the coherent case we introduce a
finite $L_\phi = \lambda$. The system is thus partitioned into phase-coherent
segments of length $L_\phi$ which are combined incoherently. For the incoherent
case, our approach assumes a vanishing phase coherence length, so no
partitioning is necessary. 

The physical system in Figure \ref{fig:circuit model} has total length
$L_\mathrm{wire} = 2\lsd + \lambda$. We will refer to this system as ``the
wire'' since, by assumption, the transport outside this region is in equilibrium
between the two spin channels and therefore does not contribute to
magnetoconductance effects. Our main interest will be the relative
magnetoconductance for the wire, $\dgongwire$, which is defined analogously to
\eqref{eq:dgongdw} by changing the intrinsic domain wall conductances to
combined wire conductances.

\subsection{Domain wall magnetoconductance in a wire: effect of disorder inside
domain wall}

\begin{figure}
\includegraphics[width=0.45\textwidth]{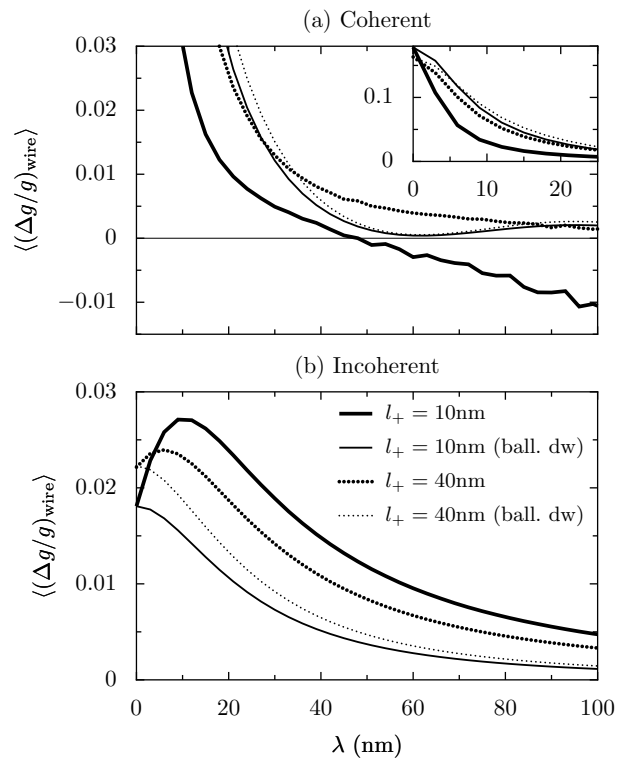}
\caption{\label{fig:dG/G vs p for circuit}
Magnetoconductance of a domain wall in a nanowire, $\langle \dgongwire \rangle$,
as a function of wall width $\lambda$ for (a) coherent and (b) incoherent
transport through the domain wall region. Two values of mean free path are
shown, $l_+ = 10\nanometer$, $40\nanometer$ (thick lines). The
magnetoconductance of an equivalent disordered wire with a ballistic domain wall
(\ie no impurities inside the wall) is also shown (thin lines). The inset in the
upper figure shows the coherent $\langle \dgongwire \rangle$ in the small
$\lambda$ region. Other parameter values are $L_y=5\nanometer$, $\rho=2$, and
$\lsd = 4l_+$, with $N_\mathrm{S} = 3000$ impurity configurations.}
\end{figure}

In Figure \ref{fig:dG/G vs p for circuit} (thick lines) we show $\langle
\dgongwire \rangle$ as a function of domain wall width $\lambda$ for both
coherent and incoherent cases. To illustrate the dependence on disorder, two
values of mean free path are shown ($l_+ = 10\nanometer, 40 \nanometer$). In all
cases we fix $\rho=2$ and $\lsd = 4l_+$; results for other values of these
parameters are qualitatively unchanged. For comparison with our previous work,
\cite{falloon2004} in which the domain wall was treated as ballistic, Figure
\ref{fig:dG/G vs p for circuit} also shows the magnetoconductance for the
equivalent wire with no scatterers inside the domain wall (thin lines). 

In the narrow wall limit, $\lambda \to 0$, the coherent regime is the physically
relevant one. Figure \ref{fig:dG/G vs p for circuit}a shows that the coherent
$\langle \dgongwire \rangle$ (solid lines) \emph{tends to a maximum value as
$\lambda \to 0$}. This is in contrast to the behaviour of the intrinsic
magnetoconductance $\langle \Delta g \rangle$ studied in the previous section,
which (for both coherent and incoherent cases) goes to zero in this limit. From
our previous work, \cite{falloon2004} this result is expected since at
$\lambda=0$ the wall corresponds to a GMR interface, \cite{valet1993} for which
there is complete mistracking of spin. It is interesting to note, however, that
for small but non-zero $\lambda$, $\langle \dgongwire \rangle$ is reduced for a
coherent wall with disorder (Figure \ref{fig:dG/G vs p for circuit}a, thick
lines) compared to a ballistic wall (Figure \ref{fig:dG/G vs p for circuit},
thin lines). Therefore, when scattering in the surrounding regions
is taken into account, \emph{the presence of disorder inside the wall reduces
the magnetoconductance effect of narrow domain walls}.

In the incoherent regime, the magnetoconductance is significantly reduced with
respect to the coherent regime in the limit $\lambda \to 0$. Furthermore, it is
a non-monotonic function of disorder in this limit, having a maximum value at a
wall width $\lambda>0$. In fact, this behaviour is an artifact arising from the
lack of complete mistracking when $\lambda\to0$ for transmission through an
incoherent wall, which we discussed in Section \ref{subsub:ballistic wall}. As
we mentioned above, in this limit the coherent regime describes the physical
situation, so these effects are not physically relevant.

In the wide wall limit $\lambda \to \infty$, the behaviour of $\langle
\dgongwire \rangle$ is qualitatively similar to that of $\langle \dgongdw
\rangle$ of Section III. The coherent magnetoconductance is significantly
reduced with respect to the incoherent one and becomes negative for large
$\lambda$ (in Figure \ref{fig:dG/G vs p for circuit}a this is most evident for
$l_+ = 10\nanometer$). These effects are due to the intrinsic properties of
coherent transport through a disordered domain wall which we discussed in
Section \ref{sub:intrinsic dg}.

For large $\lambda$ the incoherent regime is the physically relevant one. In
this case the magnetoconductance decreases as $1/\lambda^2$, which is the same
parametric behaviour as the intrinsic case (Section \ref{sub:intrinsic dgong}).
However, we will see in the next section that this effect is
larger in magnitude for $\langle \dgongwire \rangle$. In this case,
comparison between the incoherent disordered wall (Figure \ref{fig:dG/G vs p for
circuit}b, thick lines) with an incoherent ballistic wall (Figure \ref{fig:dG/G
vs p for circuit}b, thin lines) shows that \emph{the presence of
disorder inside the wall enhances the magnetoconductance effect in wide domain
walls}. This enhancement has two causes. Firstly, for a disordered wall there is
an intrinsic contribution due to scattering from impurities inside the wall,
which is absent in the case of a ballistic wall. Secondly, the disordered wall
gives rise to an enhanced spin-mistracking (Section \ref{sub:spin-dep}), which
leads to an increase in the GMR scattering in the surrounding regions.

To our knowledge no experimental data exists comparing the conductance through
domain walls in wires with different amounts of disorder; the findings of this
section suggest that such a difference may be observable. 

\subsection{Enhancement of domain wall magnetoconductance from extrinsic
scattering}

The total magnetoconductance of a wire containing a domain wall, $\dgongwire$,
arises from scattering effects which may be classified as either
\emph{intrinsic} or \emph{extrinsic}. Intrinsic effects are those due to 
scattering occurring within the region of the domain wall itself, and are
characterized by the intrinsic magnetoconductance $\dgongdw$ studied in Section
\ref{sub:intrinsic dgong}. Extrinsic effects, on the other hand, occur in the
regions adjacent to the domain wall, and arise from a GMR-like enhancement of
back-scattering due to the spin-mistracking of electrons scattered by the wall.

We now determine the relative importance of intrinsic and extrinsic effects in
the magnetoconductance of a domain wall in a nanowire. In order to do this, we
compare $\dgongwire$, which contains both intrinsic and extrinsic effects, with
$\dgongdw$, which contains only intrinsic ones.

In the narrow wall limit, $\lambda \to 0$, the amount of impurity scattering
inside the wall goes to zero, and hence $\langle \dgongdw \rangle$ becomes
negligible. On the other hand, Figure \ref{fig:dgong_int vs lambda} shows that,
in the physically relevant case of coherent walls, $\langle \dgongwire \rangle$
achieves its maximum value in this limit. It is therefore clear that \emph{for
narrow walls, extrinsic scattering effects are the dominant mechanism underlying
the domain wall magnetoconductance}.

For general $\lambda$, however, the situation is more complicated. We saw in
Section \ref{sub:intrinsic dgong} that $\langle \dgongdw \rangle$ is significant
for disordered walls. At the same time, the scattering from impurities inside
the wall enhances the spin-mistracking of transmitted and reflected electrons
(Section \ref{sub:spin-dep}), which will lead to extrinsic scattering. We
therefore expect that both intrinsic and intrinsic effects will be significant
in general. To determine the relative magnitude of the two, it is necessary
to compare $\langle \dgongwire \rangle$ and $\langle \dgongdw \rangle$
quantitatively.
 
%The coherent curves in Figure \ref{fig:dG/G vs p for circuit}a show that for
%narrow domain walls the most important contribution to magnetoconductance comes
%from an enhancement of scattering in adjacent regions due to spin-mistracking
%through the wall, in agreement with the findings of our previous work.
%\cite{falloon2004} For larger $\lambda$, we saw in Section III that the
%intrinsic scattering inside the wall becomes important, while from Figure
%\ref{fig:dG/G vs p for circuit} it can be seen that the wire magnetoconductance
%is significantly reduced from its value for small $\lambda$. For large $\lambda$
%the role of spin-dependent scattering processes in the regions adjacent to the
%wall is not clear, and in particular whether they become negligible as $\lambda
%\rightarrow \infty$. This is an important point since the most widely accepted
%existing models \cite{viret1996, levy1997, tatara1997} effectively make this
%assumption by calculating conductance only through the region of the wall. 
%
%To address this question it is therefore desirable to estimate the relative
%contribution of scattering processes which are \emph{intrinsic} (inside the wall
%region) and those which are \emph{extrinsic} (in the regions adjacent to the
%wall) as a function of the wall width $\lambda$. This can be done by comparing
%the intrinsic magnetoconductance from Section III, $\dgongdw$, with the wire
%magnetoconductance, $\dgongwire$, which contains both intrinsic and extrinsic
%contributions. 
 
For this comparison we need to ``renormalize'' $\dgongdw$ to describe the same
total system size as the wire, \ie $L_\mathrm{wire} = 2\lsd + \lambda$. To do
this, we calculate separately the \emph{total} conductances (summed over all
spin and transverse channels) for each of the three regions of the wire: the
domain wall ($g_\dw$), and the two surrounding regions ($g_\mathrm{left}$ and
$g_\mathrm{right}$). Taking the incoherent combination of the three regions
\cite{datta1997} we then obtain a total conductance $g_\mathrm{int}$ over the
length of the wire:
\begin{equation} 
\frac{1-g_\mathrm{int}}{g_\mathrm{int}} =
\frac{1-g_\mathrm{left}}{g_\mathrm{left}} +
\frac{1-g_\dw}{g_\dw} +
\frac{1-g_\mathrm{right}}{g_\mathrm{right}}. 
\end{equation}
Combining conductances in this way is equivalent to making the assumption that
the two spin channels are equilibriated at the wall boundaries rather than at a
distance $\lsd$ from the wall. 

By comparing $g_\mathrm{int}$ with the conductance of a uniform region of length
$2\lsd + \lambda$, we calculate an ``intrinsic'' magnetoconductance $\dgongint$.
This quantity represents the desired renormalization of $\dgongdw$ to the length
$L_\mathrm{wire}$. We note that for $l_+ \ll \lambda$ this can be calculated
approximately as
\begin{equation}\label{eq:renormalized intrinsic dG/G}
\dgongint = \frac{\lambda}{2\lsd+\lambda}\dgongdw.
\end{equation}

\begin{figure}
\includegraphics[width=0.45\textwidth]{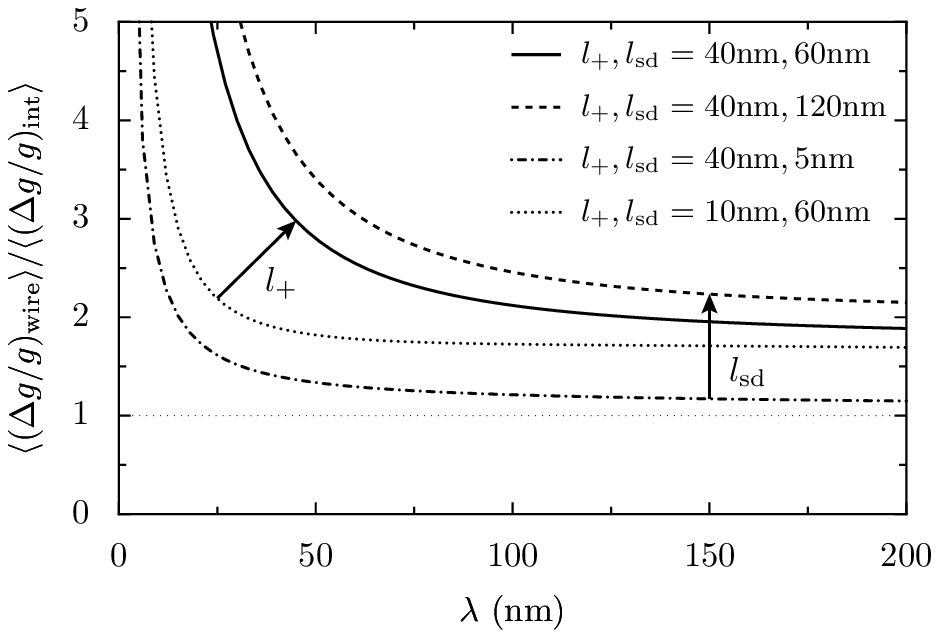}
\caption{\label{fig:dG/G circuit vs intrinsic}
Ratio of incoherent wire and intrinsic domain wall magnetoconductance $\langle
\dgongwire \rangle /\langle \dgongint \rangle$ as a function of $\lambda$, for
illustrative values of $l_+$ and $\lsd$. The system parameters are
$\rho=2$ and $L_y=10\nanometer$.}
\end{figure}

In Figure \ref{fig:dG/G circuit vs intrinsic} we show the ratio of $\langle
\dgongwire \rangle$ and $\langle \dgongint \rangle$ as a function of $\lambda$
for several values of $\rho$ and $\lsd$, in the incoherent case. We see that the
ratio is largest in the limit of small $\lambda$. As we mentioned previously,
this is because $\langle \dgongdw \rangle$ becomes negligible in the limit
$\lambda\to0$ while $\langle \dgongwire \rangle$ becomes large. We note that the
ratios shown in Figure \ref{fig:dG/G circuit vs intrinsic} apply to the
incoherent regime. As we saw in Figure \ref{fig:dG/G vs p for circuit}, $\langle
\dgongwire \rangle$ is much larger near $\lambda \to 0$ for the coherent case
than for the incoherent one. Therefore, in the coherent case the corresponding
enhancement of $\langle \dgongwire \rangle$ with respect to $\langle \dgongint
\rangle$ will also be larger.

For large $\lambda$, Figure \ref{fig:dG/G circuit vs intrinsic} shows the less
obvious result that $\langle \dgongwire \rangle$ is substantially larger than
$\langle \dgongint \rangle$ even in the case of wide walls ($\lambda \simeq 200
\nanometer$). The ratio depends on the parameter values, and for those of Figure
\ref{fig:dG/G circuit vs intrinsic} it is between $\sim$$1.2$ (small $\lsd$) and
$\sim$$2.2$ (large $\lsd$) at $\lambda = 200 \nanometer$. We thus see that
\emph{extrinsic spin-dependent scattering effects are quantitatively important
for domain wall magnetoconductance, even for wide walls}. This is an important
point, since the most widely accepted existing models \cite{viret1996, levy1997,
tatara1997} consider conductance only through the region of the wall, and hence
ignore extrinsic effects.

\section{Conclusion}

In this work we have studied spin-dependent electron transport through domain
walls in disordered ferromagnetic nanowires. Using a novel numerical scattering
matrix formalism, we considered a large range of impurity concentrations,
corresponding to transport regimes from ballistic through to diffusive. In
addition, our approach allowed us to consider spin-dependent impurity
scattering, and to distinguish between the fully coherent and fully incoherent
transport regimes.

The contribution of a domain wall to the resistance of a wire is due to two main
mechanisms. The first one is the reflection of electrons from the domain wall
itself and impurities within the domain wall region. The second is related to
the spin-dependent resistivity of the wire regions surrounding the domain wall.
We therefore considered first the spin-dependent ``intrinsic'' transport
coefficients of the disordered domain wall region itself, and then showed in the
last section what can be expected for the resistance of the whole wire, which is
the experimentally measurable quantity. 

The intrinsic effects stem from the presence of impurities inside a domain wall,
which causes electrons to be scattered both with and without spin reversal. This
leads to an enhancement of backscattering caused by exposure of electrons in a
superposition of spin states to spin-dependent scattering. In the incoherent
regime this gives rise to a reduction of the intrinsic domain wall conductance.
The maximum reduction occurs at a disorder-dependent value of the domain wall
length. 
 
In the phase-coherent regime, the relative amount of reflection with spin
reversal is reduced with respect to the incoherent regime. For large disorder,
this acts to suppress the backscattering enhancement observed in the incoherent
case. Indeed, the overall effect in this regime is an enhancement of
\emph{transmission} through the domain wall. This can be understood as arising
from a suppression of weak localization by the domain wall, which is consistent
with the findings of Ref.~\onlinecite{tatara1997}. However, we point out several
reasons why such a behaviour would be difficult to observe experimentally.

The mesoscopic character of the transport in the coherent regime is underlined
by a study of the conductance fluctuations. For a domain wall it is found that
these are reduced with respect to the uniformly magnetized case, as a result of
decorrelation between the components of a given incoming spin state which are
transmitted with and without reversal of spin. Furthermore, the sample-dependent
fluctuations of the domain wall contribution to the resistance can be
significantly larger than its average value in the diffusive limit. This should
be kept in mind when experiments on individual samples approach the coherent
regime. 

To address the experimentally measurable conductance of a disordered
nanowire containing a domain wall, we included in our treatment uniformly
magnetized regions of length $\lsd$ (the spin diffusion length) on either side
of the domain wall. Since the phase coherence length does not exceed $\lsd$, the
rest of the wire is taken into account as a classical resistance in series that
does not contribute to the magnetoconductance.

The spin-dependent scattering in the regions adjacent to the wall results in an
enhancement of the resistance. This arises from processes in which the spin of
the electrons does not follow the domain wall magnetization while they propagate
through the wall. The spin-dependent scattering in the surrounding regions
contributes an ``extrinsic'' domain wall resistance effect which is similar in
nature to giant magnetoresistance. This mechanism is distinct from the intrinsic
effect occurring inside the wall.

As compared to the intrinsic domain wall resistance alone, the total resistance
due to a domain wall in a nanowire is significantly enhanced by the extrinsic
scattering. The enhancement is most dramatic for narrow walls, where the
intrinsic resistance becomes negligible. However, it remains important (a factor
of $2$ for typical parameters) even for wide walls. 

In our previous study based on ballistic domain walls in a circuit model
\cite{falloon2004} we found results in order-of-magnitude agreement with a
recent experiment. \cite{ebels2000} The relevant regime for this experiment is
intermediate between the coherent and incoherent regimes, and corresponds to
domain walls far from the adiabatic limit of zero spin-mistracking. In the
present work, we have found that the presence of disorder inside the wall in
this regime leads to a quantitative reduction of the magnetoconductance effect
as compared to the ballistic wall used in Ref.~\onlinecite{falloon2004}, but is
still within an order of magnitude of the experiment. This is acceptable given
the uncertainties over various physical parameters relevant for the experiment
(such as the precise value of the elastic mean free path). On the other hand, in the
case of wide walls in the incoherent transport regime, we have found that
disorder enhances the magnetoconductance effect. This regime is relevant for
experiments in materials such as nickel and iron. Our model should also be
applicable to the understanding of recent measurements of domain wall resistance
in ferromagnetic semiconductors. \cite{chiba2006} 

In both narrow and wide wall regimes, our results show that disorder inside and
outside a domain wall leads to important effects in the magnetoconductance.
Since disorder is unavoidable in mesoscopic wires, our conclusions should be
testable experimentally.

\begin{acknowledgments} 
PEF is grateful for the support of an Australian Postgraduate Award and a Jean
Rogerson Fellowship from the University of Western Australia and for support
from the Universit\'e Louis Pasteur in Strasbourg. This work was also supported
through the Australian Research Council Linkages and Discovery Programmes and by
the European Union through the RTN programme. RAJ and DW acknowledge the
hospitality the Physics Department of the University of Western Australia, where
part of this work was done with financial support from the ARC (RAJ) and the
DREI of the CNRS (DW).
\end{acknowledgments}

\appendix

\section{Scattering matrix of domain wall interfaces}
\label{app:dw S matrix}

In this appendix we present formulas for the calculation of the scattering
matrices of the domain wall interfaces $\tilde{s}^{(\mathrm{L,R})}$. These
are obtained by forming scattering state solutions from the basis states
of Eqs.~(\ref{eq:plain basis states}) and (\ref{eq:spiral basis state}) for each
side of the interface and matching the functions and their derivatives at $z=0$
(left interface) or $z=\lambda$ (right interface). For the domain wall
interfaces the matrix elements corresponding to scattering between different
transverse modes, $n' \neq n$, are zero since the domain wall profile depends
only on the longitudinal coordinate. The amplitudes $\tilde{r} ^{\mathrm{(L,R)}}
_{\pm \sigma n; \sigma n}$, $\tilde{t} ^\mathrm{(L,R)} _{\pm \sigma n; \sigma
n}$ and $\tilde{r}'^{\mathrm{(L,R)}}_{\pm \sigma n; \sigma n}$,
$\tilde{t}'^{\mathrm{(L,R)}}_{\pm \sigma n; \sigma n}$ satisfy $4\times4$ sets
of equations. The equations for $\tilde{r} ^{(\mathrm{L})} _{\sigma'n; \sigma
n}$ and $\tilde{t} ^{(\mathrm{L})} _{\sigma'n; \sigma n}$ read (for clarity we
omit the $n$ and $\mathrm{L}$ labels)
\begin{widetext}
\begin{subequations} \label{eq:left interface matching left}
\begin{eqnarray}
1+\tilde{r}_{\sigma\sigma} &=& 
\sqrt{\frac{v_{\sigma}}{\tilde{v}_{\sigma}}} \tilde{t}_{\sigma\sigma}
+ \rmi A_{-\sigma}\sqrt{\frac{v_{\sigma}}{\tilde{v}_{-\sigma}}}
\tilde{t}_{-\sigma\sigma}, \\
\sqrt{\frac{v_{\sigma}}{v_{-\sigma}}} \tilde{r}_{-\sigma\sigma}
&=& \rmi A_{\sigma} \sqrt{\frac{v_{\sigma}}{\tilde{v}_{\sigma}}}
\tilde{t}_{\sigma\sigma}
+ \sqrt{\frac{v_{\sigma}}{\tilde{v}_{-\sigma}}}
\tilde{t}_{-\sigma\sigma}, \\
\rmi k_{\sigma} \left(1 - \tilde{r}_{\sigma\sigma}\right)
&=& \rmi\left(\tilde{k}_{\sigma} - \sigma k_\lambda A_{\sigma}\right)
\sqrt{\frac{v_{\sigma}}{\tilde{v}_{\sigma}}} \tilde{t}_{\sigma\sigma}
- \left(\sigma k_\lambda + \tilde{k}_{-\sigma}A_{-\sigma}\right)
\sqrt{\frac{v_{\sigma}}{\tilde{v}_{-\sigma}}}
\tilde{t}_{-\sigma\sigma}, \\
-\rmi k_{-\sigma}\sqrt{\frac{v_{\sigma}}{v_{-\sigma}}}
\tilde{r}_{-\sigma\sigma}
&=& \left(\sigma k_\lambda - \tilde{k}_{\sigma}A_{\sigma}\right)
\sqrt{\frac{v_{\sigma}}{\tilde{v}_{\sigma}}} \tilde{t}_{\sigma\sigma}
+ \rmi\left(\tilde{k}_{-\sigma}+\sigma k_\lambda A_{-\sigma}\right)
\sqrt{\frac{v_{\sigma}}{\tilde{v}_{-\sigma}}}
\tilde{t}_{-\sigma\sigma},
\end{eqnarray}
\end{subequations}
while for $\tilde{r}'^{(\mathrm{L})}_{\sigma'n; \sigma n}$ and
$\tilde{t}'^{(\mathrm{L})}_{\sigma'n; \sigma n}$ we have
\begin{subequations} \label{eq:left interface matching right}
\begin{eqnarray}
1 + \tilde{r}_{\sigma \sigma}'
+ \rmi A_{-\sigma} \sqrt{\frac{\tilde{v}_{\sigma}}{\tilde{v}_{-\sigma}}}
\tilde{r}_{-\sigma \sigma}'
&=& \sqrt{\frac{\tilde{v}_{\sigma}}{v_{\sigma}}} 
\tilde{t}_{\sigma \sigma}', \\
-\rmi A_{\sigma}\left(1 - \tilde{r}_{\sigma \sigma}'\right)
+\sqrt{\frac{\tilde{v}_{\sigma}}{\tilde{v}_{-\sigma}}}
\tilde{r}_{-\sigma \sigma}' 
&=& \sqrt{\frac{\tilde{v}_{\sigma}}{v_{-\sigma}}}
\tilde{t}_{-\sigma \sigma}', \\
-\rmi\left(\tilde{k}_{\sigma} - \sigma k_\lambda A_{\sigma}\right)
\left(1 - \tilde{r}_{\sigma \sigma}'\right)
- \left(\sigma k_\lambda + \tilde{k}_{-\sigma} A_{-\sigma}\right)
\sqrt{\frac{\tilde{v}_{\sigma}}{v_{-\sigma}}}
\tilde{r}_{-\sigma \sigma}'
&=& -\rmi k_{\sigma} \sqrt{\frac{\tilde{v}_{\sigma}}{v_{\sigma}}}
\tilde{t}_{\sigma\sigma}', \\
\left(\sigma k_\lambda - \tilde{k}_{\sigma}A_{\sigma}\right)
\left(1 + \tilde{r}_{\sigma \sigma}'\right)
+\rmi \left(\tilde{k}_{-\sigma} + \sigma k_\lambda A_{-\sigma}\right)
\sqrt{\frac{\tilde{v}_{\sigma}}{v_{-\sigma}}}
\tilde{r}_{-\sigma \sigma}'
&=& -\rmi k_{-\sigma} \sqrt{\frac{\tilde{v}_{\sigma}}{v_{-\sigma}}}
\tilde{t}_{-\sigma \sigma}'.
\end{eqnarray}
\end{subequations}
\end{widetext}
We present only the equations for the elements of $\tilde{s} ^{(\mathrm{L})}$
explicitly; because of the symmetry of the spiral domain wall profile
(\eqref{eq:domain wall profile}), the elements of $\tilde{s} ^{(\mathrm{R})}$
can be found from Eqs.~(\ref{eq:left interface matching left}--\ref{eq:left
interface matching right}) after interchanging $r_{\sigma'\sigma}
\leftrightarrow r_{\sigma'\sigma}'$, $t_{\sigma'\sigma} \leftrightarrow
t_{\sigma'\sigma}'$ and replacing $k_\lambda \to -k_\lambda$. 

\section{Delta function scattering matrices in uniform and domain
wall potentials}
\label{app:delta function s matrix}

We now consider the scattering matrices for a delta function scatterer located
in regions of uniform and rotating magnetization, $s^{(\delta)}(y_\alpha)$ and
$\tilde{s}^{(\delta)}(y_\alpha)$. In the uniform case, $s^{(\delta)}(y_\alpha)$
can be obtained using the result for a spin-independent delta function
potential. \cite{cahay1988} The individual amplitudes are given by
\begin{subequations}
\label{eq:delta fn solution uniform}
\begin{eqnarray}
t^{\mathrm{(\delta)}}_{\sigma'n';\sigma n} &=&
t'^{\mathrm{(\delta)}}_{\sigma'n';\sigma n} =
\delta_{\sigma'\sigma}
\left[\mathcal{M}_\sigma\right]_{n'n}^{-1},\\
r^{\mathrm{(\delta)}}_{\sigma'n';\sigma n} &=&
r'^{\mathrm{(\delta)}}_{\sigma'n';\sigma n} =
t^{\mathrm{(\delta)}}_{\sigma'n';\sigma n} - \delta_{n'n},
\end{eqnarray}
\end{subequations}
where $\mathcal{M}_\sigma$ is an $N_\sigma \times N_\sigma$ matrix with elements
\begin{equation}
\left[\mathcal{M}_\sigma\right]_{n'n} = 
\delta_{n'n} + \frac{\rmi u_\sigma\phi_{n'}(y_\alpha)\phi_n(y_\alpha)}
                   {\hbar\sqrt{v_{\sigma n'}v_{\sigma n}}}.
\end{equation}

The scattering matrix for a delta function located inside a domain wall,
$\tilde{s}^{(\delta)}(y_\alpha)$, is more complicated because the domain wall
basis states $\tilde{\psi}_{\sigma n}(\vec{r})$ couple up and down components.
Defining $\beta_n = A_{\sigma n}A_{-\sigma n}$ we have
\begin{widetext}
\begin{subequations}
\label{eq:delta fn spiral s matrix}
\begin{alignat}{3}
\tilde{t}^{(\delta)}_{\sigma n';\sigma n} & =
\tilde{t}'^{(\delta)}_{\sigma n';\sigma n} && = 
\sqrt{\frac{\tilde{v}_{\sigma n'}}{\tilde{v}_{\sigma n}}}
\frac{1}{1+\beta_{n'}}\left(
\left[\tau_{\sigma \sigma}\right]_{n'n} - \rmi A_{-\sigma n'} 
\left[\tau_{-\sigma \sigma}\right]_{n'n} \right), \\
\tilde{t}^{(\delta)}_{-\sigma n';\sigma n} & =
-\tilde{t}'^{(\delta)}_{-\sigma n';\sigma n} && =
\sqrt{\frac{\tilde{v}_{-\sigma n'}}{\tilde{v}_{\sigma n}}}
\frac{1}{1+\beta_{n'}}\left(
-\rmi A_{\sigma n'}\left[\tau_{\sigma \sigma}\right]_{n'n} 
+ \left[\tau_{-\sigma \sigma}\right]_{n'n} \right), \\
\tilde{r}^{(\delta)}_{\sigma n'; \sigma n} & =
\tilde{r}'^{(\delta)}_{\sigma n';\sigma n} && = 
\sqrt{\frac{\tilde{v}_{\sigma n'}}{\tilde{v}_{\sigma n}}}
\frac{1}{1+\beta_{n'}}\left(
\left[\tau_{\sigma \sigma}\right]_{n'n} 
+ \rmi A_{-\sigma n'}\left[\tau_{-\sigma \sigma}\right]_{n'n}
- (1-\beta_{n'})\delta_{n'n} \right), \\
\tilde{r}^{(\delta)}_{-\sigma n'; \sigma n} & =
-\tilde{r}'^{(\delta)}_{-\sigma n';\sigma n} && =
\sqrt{\frac{\tilde{v}_{-\sigma n'}}{\tilde{v}_{\sigma n}}}
\frac{1}{1+\beta_{n'}}\left(
\rmi A_{\sigma n'}\left[\tau_{\sigma \sigma}\right]_{n'n} 
+ \left[\tau_{-\sigma \sigma}\right]_{n'n} 
- 2\rmi A_{\sigma n'} \delta_{n'n} \right),
\end{alignat}
\end{subequations}
where $\tau_{\sigma'\sigma}$ and $\tilde{\mathcal{M}}_\sigma$ are matrices of
dimension $N_{\sigma'}\times N_\sigma$ and $N_\sigma\times N_\sigma$
respectively, defined by
\begin{subequations}
\begin{eqnarray}
\left[\tau_{\sigma \sigma}\right]_{n'n} &=& 
\left[\tilde{\mathcal{M}}_\sigma^{-1}\right]_{n'n},\\
\left[\tau_{-\sigma \sigma}\right]_{n'n} &=& 
\rmi A_{\sigma n} \left[\tilde{\mathcal{M}}_{-\sigma}^{-1}\right]_{n'n},\\
\left[\tilde{\mathcal{M}}_\sigma\right]_{n'n} &=& \delta_{n'n} -
\frac{mu_\sigma}{\rmi\hbar^2}\frac{(1+\beta_{n'}) \phi_{n'}(y_\alpha)\phi_n(y_\alpha)}
{\tilde{k}_{\sigma n'} + \beta_{n'}\tilde{k}_{-\sigma n'}}.
\end{eqnarray}
\end{subequations}
%\end{widetext}

The amplitudes in Eqs.~(\ref{eq:delta fn solution uniform}) and (\ref{eq:delta
fn spiral s matrix}) determine the matrices $s^{(\delta)}(y_\alpha)$ and
$\tilde{s}^{(\delta)}(y_\alpha)$ that are composed numerically for each impurity
configuration in order to obtain the total transmission coefficients in various
regimes.

\section{Delta function scattering in weak disorder limit}
\label{app:Born approximation}

In the limit of weak disorder, $u_\sigma \rightarrow 0$, transport through the
impurity potential $V(\vec{r})$ (Eqs.~(\ref{eq:Vsigma}--\ref{eq:V def2})) can be
treated within the Born approximation. This approach yields the scattering
amplitudes to lowest order in $u_\sigma$. For the uniformly magnetized case, we
have
%
%\begin{widetext}
%
%\begin{subequations} \label{eq:delta born uniform}
%\begin{eqnarray}
%%
%r^{(\delta)}_{\sigma'n';\sigma n} &=&
%r'^{(\delta)}_{\sigma'n';\sigma n} =
%-\delta_{\sigma'\sigma}
%\frac{\rmi u_\sigma\phi_{n'}(y_\alpha)\phi_n(y_\alpha)}
%     {\hbar\sqrt{v_{\sigma n'} v_{\sigma n}}},\\
%\label{eq:delta born uniform r}
%%
%t^{(\delta)}_{\sigma'n';\sigma n} &=&
%t'^{(\delta)}_{\sigma'n';\sigma n} =
%\delta_{\sigma'\sigma} \delta_{n'n} + r^{(\delta)}_{\sigma' n';\sigma n},
%\label{eq:delta born uniform t}
%%
%\end{eqnarray}
%\end{subequations}
%
%
\begin{subequations} \label{eq:delta born uniform}
\begin{eqnarray}
t^{(\uni)}_{\sigma'n';\sigma n} &=&
\delta_{\sigma'\sigma} \rme^{\rmi k_{\sigma'n'}\lambda}
\left\{\delta_{n'n} -
\frac{\rmi u_\sigma}{\hbar\sqrt{v_{\sigma n'} v_{\sigma n}}}
\sum_{\alpha=1}^{\Nimp}\phi_{n'}(y_\alpha)\phi_n(y_\alpha)
\rme^{\rmi(k_{\sigma n}-k_{\sigma n'})x_\alpha}
\right\} + O(u_\sigma^2), \\
r^{(\uni)}_{\sigma'n';\sigma n} &=&
-\delta_{\sigma'\sigma}
\frac{\rmi u_\sigma}{\hbar\sqrt{v_{\sigma n'} v_{\sigma n}}}
\sum_{\alpha=1}^{\Nimp}\phi_{n'}(y_\alpha)\phi_n(y_\alpha)
\rme^{\rmi(k_{\sigma n'}+k_{\sigma n})x_\alpha} + O(u_\sigma^2).
\end{eqnarray}
\end{subequations}
\end{widetext}

In the case of a domain wall, the situation is more complicated because
scattering from the interfaces at $z=0, \lambda$ leads to mixing of up and down
spin channels. It is most convenient to first calculate amplitudes for transport
through the disordered region $0<z<\lambda$, not including the interfaces, which
we write $\tilde{t}^{(\Nimp \times \delta)} _{\sigma'n'; \sigma n}$ and
$\tilde{r} ^{(\Nimp \times \delta)} _{\sigma'n'; \sigma n}$. Formally, these
coefficients correspond to a scenario in which the rotating potential $\theta(z)
= \pi z/\lambda$ is valid for \emph{all} $z$, so that the asymptotic states are
the domain wall basis states $\tilde{\psi}_{\sigma n}^\gtrless(\vec{r})$. Within
the Born approximation, these amplitudes are given by
\begin{widetext}
\begin{subequations}\label{eq:delta born spiral}
\begin{eqnarray}
\tilde{t}^{(\Nimp\times\delta)}_{\sigma n';\sigma n} &=&
\rme^{\rmi\tilde{k}_{\sigma n'}\lambda}
\left\{\delta_{n'n} 
-\frac{\rmi\left(u_\sigma + u_{-\sigma} A_{\sigma n'}A_{\sigma n}\right)}
      {\hbar\sqrt{\tilde{v}_{\sigma n'}\tilde{v}_{\sigma n}}}
\sum_{\alpha=1}^{\Nimp}
\phi_{n'}(y_\alpha)\phi_n(y_\alpha)
\rme^{\rmi(\tilde{k}_{\sigma n}-\tilde{k}_{\sigma n'})x_\alpha}
\right\} + O(u_\sigma^2),\\
\tilde{t}^{(\Nimp\times\delta)}_{-\sigma n';\sigma n} &=&
\rme^{\rmi\tilde{k}_{-\sigma n'}\lambda}
\frac{u_{-\sigma} A_{\sigma n} - u_\sigma A_{-\sigma n'}}
     {\hbar\sqrt{\tilde{v}_{-\sigma n'}\tilde{v}_{\sigma n}}}
\sum_{\alpha=1}^{\Nimp}
\phi_{n'}(y_\alpha)\phi_n(y_\alpha)
\rme^{\rmi(\tilde{k}_{\sigma n}-\tilde{k}_{-\sigma n'})x_\alpha} 
+ O(u_\sigma^2), \\
\tilde{r}^{(\Nimp\times\delta)}_{\sigma n';\sigma n} &=&
-\frac{\rmi\left(u_\sigma-u_{-\sigma}A_{\sigma n'}A_{\sigma n}\right)}
      {\hbar\sqrt{\tilde{v}_{\sigma n'}\tilde{v}_{\sigma n}}}
\sum_{\alpha=1}^{\Nimp}
\phi_{n'}(y_\alpha)\phi_n(y_\alpha)
\rme^{\rmi(\tilde{k}_{\sigma n}+\tilde{k}_{\sigma n'})x_\alpha}
+ O(u_\sigma^2), \\
\tilde{r}^{(\Nimp\times\delta)}_{-\sigma n';\sigma n} &=&
\frac{u_{-\sigma} A_{\sigma n} + u_\sigma A_{-\sigma n'}}
      {\hbar\sqrt{\tilde{v}_{\sigma n'}\tilde{v}_{\sigma n}}}
\sum_{\alpha=1}^{\Nimp}
\phi_{n'}(y_\alpha)\phi_n(y_\alpha)
\rme^{\rmi(\tilde{k}_{\sigma n}+\tilde{k}_{-\sigma n'})x_\alpha} 
+ O(u_\sigma^2).
\end{eqnarray}
\end{subequations}
\end{widetext}

To calculate the \emph{average} scattering probabilities, it is necessary to
take the squared magnitude of the quantities in Eqs.~(\ref{eq:delta born
uniform}--\ref{eq:delta born spiral}) and then average over the impurity
positions $\vec{r}_\alpha$. Because the positions of different impurities are
uncorrelated, the average of terms involving two different impurities are a
factor $1/\kF L_z$ smaller than those involving a single impurity. They are
therefore negligible, and the scattering is dominated by the single impurity
scattering events. \cite{jalabert1995}

We note that the probabilities corresponding to the amplitudes
$t^{(\uni)}_{\sigma n;\sigma n}$ and $\tilde{t}^{(\Nimp \times \delta)} _{\sigma
n;\sigma n}$, which are equal to unity in the limit $u_\sigma = 0$, cannot be
calculated by simply taking the absolute square of the amplitudes in
Eqs.~(\ref{eq:delta born uniform}a) and (\ref{eq:delta born spiral}a). This is
due to presence of the factor $\delta_{n'n}$, which means that the term of
$O(u_\sigma^2)$ in the amplitude (which is not given in Eqs.~(\ref{eq:delta born
uniform}a) and (\ref{eq:delta born spiral}a)) also contributes a term of
$O(u_\sigma^2)$ in the corresponding probability. It is therefore necessary
either to calculate the amplitudes to $O(u_\sigma^2)$ or, more conveniently, to
use conservation of probability to express it in terms of the other
probabilities.

With these considerations, we find the average probabilities in the uniform case
as follows:
\begin{subequations} \label{eq:born uniform average}
\begin{eqnarray}
\langle R^{(\uni)}_{\sigma n';\sigma n} \rangle &=&
\nimp u_\sigma^2\frac{L_z}{L_y}
\frac{\left( 1 + \frac{1}{2} \delta_{n'n} \right)}
     { \hbar^2 v_{\sigma n'} v_{\sigma n} }, \\
\langle T^{(\uni)}_{\sigma n';\sigma n} \rangle &=&
\langle R^{(\uni)}_{\sigma n';\sigma n} \rangle, \quad n'\neq n, \\
\langle T^{(\uni)}_{\sigma n;\sigma n} \rangle &=&
1 - \nimp u_\sigma^2 \frac{L_z}{L_y}
\sum_{n'=1}^{N_\sigma} \frac{2-\frac{1}{2}\delta_{n'n}}
                            {\hbar^2v_{\sigma n'}v_{\sigma n}}.
\end{eqnarray}
\end{subequations}

Summing over all $n$ and $n'$, we can express the total spin-dependent
transmission and reflection in terms of the mean free path $l_\sigma$
(\eqref{eq:mfp 2D}):
\begin{subequations} \label{eq:born total RT spin-dep uniform}
\begin{eqnarray}
\label{eq:born total R spin-dep uniform}
\langle R_{\sigma'\sigma}^{(\uni)} \rangle &=& \delta_{\sigma'\sigma}
\frac{\pi}{2} \frac{N_\sigma L_z}{l_\sigma} F_1(N_\sigma), \\
\label{eq:born total T spin-dep uniform}
\langle T_{\sigma'\sigma}^{(\uni)} \rangle &=& N_\sigma -
\langle R_{\sigma'\sigma}^{(\uni)} \rangle,
\end{eqnarray}
\end{subequations}
where
\begin{equation}
F_1(N_\sigma) = \frac{k_{\sigma,\mathrm{F}}L_y}{\pi N_\sigma}
\left(\frac{2}{L_y}\right)^2
\sum_{n'=1}^{N_\sigma}\sum_{n=1}^{N_\sigma}
\frac{1+\frac{1}{2}\delta_{n'n}}{k_{\sigma n}k_{\sigma n'}}
\,.
\end{equation}
Here $F_1(N_\sigma)$ is a form factor which takes account of the finite number
of channels and reduces to unity in the limit $N_\sigma \to \infty$. 

In the domain wall case, the probabilities corresponding to the amplitudes in 
\eqsref{eq:delta born spiral} are given by
\begin{widetext}
\begin{subequations}
\label{eq:inside dw RT average}
\begin{eqnarray}
\langle \tilde{R}^{(\Nimp\times\delta)}_{\sigma n';\sigma n} \rangle &=&
\frac{\nimp L_z}{L_y} \frac{1+\frac{1}{2}\delta_{n'n}}
                        {\hbar^2 \tilde{v}_{\sigma n'}\tilde{v}_{\sigma n}}
\left(u_\sigma - u_{-\sigma} A_{\sigma n'}A_{\sigma n} \right)^2, \\
\langle \tilde{R}^{(\Nimp\times\delta)}_{-\sigma n';\sigma n} \rangle &=&
\frac{\nimp L_z}{L_y} \frac{1+\frac{1}{2}\delta_{n'n}}
                        {\hbar^2 \tilde{v}_{-\sigma n'}\tilde{v}_{\sigma n}}
\left(u_{-\sigma}A_{\sigma n} + u_\sigma A_{-\sigma n'} \right)^2, \\
\langle \tilde{T}^{(\Nimp\times\delta)}_{\sigma n';\sigma n} \rangle &=&
\frac{\nimp L_z}{L_y} \frac{1+\frac{1}{2}\delta_{n'n}}
                        {\hbar^2 \tilde{v}_{\sigma n'}\tilde{v}_{\sigma n}}
\left(u_\sigma + u_{-\sigma} A_{\sigma n'}A_{\sigma n} \right)^2,
\quad n'\neq n, \\
\langle \tilde{T}^{(\Nimp\times\delta)}_{-\sigma n';\sigma n} \rangle &=&
\frac{\nimp L_z}{L_y} \frac{1+\frac{1}{2}\delta_{n'n}}
                        {\hbar^2 \tilde{v}_{-\sigma n'}\tilde{v}_{\sigma n}}
\left(u_{-\sigma}A_{\sigma n} - u_\sigma A_{-\sigma n'} \right)^2, \\
\langle \tilde{T}^{(\Nimp\times\delta)}_{\sigma n;\sigma n} \rangle &=&
1 - 
\langle \tilde{R}^{(\Nimp\times\delta)}_{\sigma n;\sigma n} \rangle 
- \sum_{\substack{n'=1, \\ n'\neq n}}^{\tilde{N}_{\sigma}}
\left[
\langle \tilde{R}^{(\Nimp\times\delta)}_{\sigma n';\sigma n} \rangle +
\langle \tilde{T}^{(\Nimp\times\delta)}_{\sigma n';\sigma n} \rangle 
\right]
- \sum_{n'=1}^{\tilde{N}_{-\sigma}} 
\left[
\langle \tilde{R}^{(\Nimp\times\delta)}_{-\sigma n';\sigma n} \rangle +
\langle \tilde{T}^{(\Nimp\times\delta)}_{-\sigma n';\sigma n} \rangle 
\right].
\end{eqnarray}
\end{subequations}
%
%\end{widetext}

In general, it is non-trivial to incorporate the interfaces and thus obtain the
total transmission probabilities for the domain wall, since if there is non-zero
reflection from the interfaces then multiple paths through the disordered region
need to be considered. The situation simplifies somewhat in the wide wall limit
of \eqsref{eq:interface asymptotic}, since to $O(1/p_\mathrm{F}^2)$ the
interface reflection probabilities are zero. The transmission through the entire
domain wall, including both interfaces and the disordered region, can then be
found by multiplying the transmission of each part. Furthermore, since the
interface transmission is diagonal in transverse channel number, only
intermediate spin states need to be summed over.

It is simplest to consider the incoherent case, since the probabilities are
taken before combining with the interfaces, meaning that the results of
Eqs.~(\ref{eq:born uniform average}--\ref{eq:inside dw RT average}) can be used
directly. In particular, for the transmission and reflection probabilities from
the left we have
%
%\begin{widetext}
%
\begin{subequations}
\begin{eqnarray}
\langle\tilde{T}^{(\dw)}_{\sigma'n';\sigma n}\rangle &=&
\sum_{\sigma_1,\sigma_2 = \pm}
\tilde{T}^{(\mathrm{R})}_{\sigma'n';\sigma_1 n'}
\langle\tilde{T}^{(\Nimp\times\delta)}_{\sigma_1n';\sigma_2n}\rangle
\tilde{T}^{(\mathrm{L})}_{\sigma_2n;\sigma n}\,, \\
\langle\tilde{R}^{(\dw)}_{\sigma'n';\sigma n}\rangle &=&
\sum_{\sigma_1,\sigma_2 = \pm}
\tilde{T}'^{(\mathrm{L})}_{\sigma'n';\sigma_1 n'}
\langle\tilde{R}^{(\Nimp\times\delta)}_{\sigma_1n';\sigma_2n}\rangle
\tilde{T}^{(\mathrm{L})}_{\sigma_2n;\sigma n}\,.
\end{eqnarray}
\end{subequations}

The total spin-dependent probabilities, $\langle \tilde{T} ^{(\dw)} _{\sigma'
\sigma} \rangle$ and $\langle \tilde{R} ^{(\dw)} _{\sigma' \sigma} \rangle$, may
be found by summing over the previous expressions, yielding quite cumbersome
expressions. Of particular interest, however, is the transmission with
spin-mistracking, $\langle \tilde{T} ^{(\dw)} _{-\sigma \sigma} \rangle$, which
is given by
\begin{equation} \label{eq:born tmp total}
\langle\tilde{T}^{(\dw)}_{-\sigma\sigma}\rangle =
\frac{2N_\sigma}{3\pF^2} 
\left\{F_2(N_\sigma) -
  \frac{6}{\pi}\frac{L_z}{l_+}\frac{N_\sigma}{\kF L_y/\pi} 
  \left[
    \rho + \frac{\pi^2}{8}(1+\rho^2)F_3(N_\sigma) +
    \frac{(1+\rho)^2}{8N_\sigma}
  \right] 
\right\} \,, 
\end{equation}
where
\begin{eqnarray}
F_2(N_\sigma) &=& \frac{3}{2}\left[
1 - \left(\frac{\pi}{\kF L_y}\right)^2
\frac{(N_\sigma+1)(2N_\sigma+1)}{6}
\right], \\
F_3(N_\sigma) &=& \frac{8}{N_\sigma^2\pi^2}
\sum_{n'=1}^{N_\sigma}\sum_{n=1}^{N_\sigma}
\frac{k_{0n}}{k_{0n'}}\,.
\end{eqnarray}
Here $F_2(N_\sigma)$ and $F_3(N_\sigma)$ are form factors which reduce to unity
in the limit $L_y \to \infty$. Note that in the approximation used here we have
$k_{+,\mathrm{F}} = k_{-,\mathrm{F}} = \kF$ and $N_+ = N_-$. \eqref{eq:born tmp
total} shows that the initial slope of $\tilde{T}^{(\dw)}_{-\sigma\sigma}$ as a
function of $1/l_+$ is \emph{always negative}, which explains the observed
behaviour the inset of Figure \ref{fig:dw tr spin-dep}.

Another important insight permitted by the perturbative approach is the
spin-dependent difference in transmission $\Delta T_\pm$. Here we find
\begin{equation} \label{eq:delta T born}
\langle \Delta T_\sigma \rangle =
-\sigma \frac{4}{\pi}\frac{\rho-1}{\pF^2}\frac{N_\sigma L_z}{l_+}
\left(\frac{N_\sigma\pi}{\kF L_y}\right)
\left\{
\frac{\pi^2}{8}F_3(N_\sigma) + \frac{1}{2N_\sigma}
\right\} \,.
\end{equation}
\end{widetext}
\eqref{eq:delta T born} shows that $\langle \Delta T_\sigma \rangle$ increases
linearly with $1/l_+$ for weak disorder, with a slope whose sign depends on
$\sigma$. This explains the behaviour of $\langle \Delta T_\sigma \rangle$
observed in Figure \ref{fig:delta G vs disorder}b. 

Interestingly, the sum $\langle \Delta T_+\rangle + \langle\Delta T_- \rangle$
in \eqref{eq:delta T born} is zero, implying that the intrinsic
magnetoconductance $\langle \Delta g\rangle$ is zero to first order in $1/l_+$.
This is in agreement with our numerical results for $\langle\Delta g\rangle$,
which show a quadratic dependence on $1/l_+$ for small disorder.

Finally, we point out that for spin-independent disorder, $\rho=1$,
\eqref{eq:delta T born} predicts that $\langle \Delta T_\sigma \rangle = 0$.
This is because in this case, the only spin-dependence of scattering comes from
the difference in the up and down wavevectors, which is treated as zero in the
approximation of \eqsref{eq:interface asymptotic}.

\section{Spin-mixing in a one-dimensional model}
\label{app:spin-mixing 1D}
The results of Section \ref{sub:spin-dep} for the transmission and reflection
through a sequence of impurity scatterings show that repeated scattering from
delta functions in a domain wall leads to a relative increase of the spin-mixing
transmission and reflection coefficients, $T_{-\sigma\sigma}$ and
$R_{-\sigma\sigma}$. We now gain some insight into this mechanism through a
simple toy model which captures some essential features of the problem. In
particular, consider a one-dimensional sequence of idealized spin-mixing
scatterers for which there is zero reflection and for which the spin-dependent
transmission probabilities are
\begin{equation}
T = T' = \mymatrix{1-\epsilon_+}{\epsilon_-}{\epsilon_+}{1-\epsilon_-}.
\end{equation}
Here $\epsilon_\pm$ represent the off-diagonal scattering. Technically, this
model is actually an example of a \emph{Markov chain}, widely studied in
probability theory. \cite{grimmett2004}

The absence of reflection allows us to calculate the total transmission through
$\Nimp$ scatterers by taking $T^{\Nimp}$. Diagonalizing $T$ and taking powers of
the diagonal components, we evaluate this as
\begin{equation} \label{eq:toy T spin-dep}
T^{\Nimp} = \frac{1}{\epsilon_++\epsilon_-}
\mymatrix{\epsilon_- + \epsilon_+\chi^{\Nimp}}
{\epsilon_-\left(1-\chi^{\Nimp}\right)}
{\epsilon_+\left(1-\chi^{\Nimp}\right)}
{\epsilon_+ + \epsilon_-\chi^{\Nimp}},
\end{equation}
where $\chi=1-\epsilon_+-\epsilon_-$. In the limit of infinitely many scatterers
this reduces to
\begin{equation}\label{eq:toy T spin-dep limit}
T^{\infty} = \frac{1}{\epsilon_++\epsilon_-}
\mymatrix{\epsilon_-}{\epsilon_-}{\epsilon_+}{\epsilon_+}.
\end{equation}
%
%In the special case of spin-independent disorder, $\epsilon_+ = \epsilon_- =
%\epsilon$, we have
%%
%\begin{subequations}
%\begin{eqnarray}\label{eq:ideal T limit spin-indep}
%T^{\Nimp} &=& \frac{1}{2}
%\mymatrix{1+\left(1-2\epsilon\right)^{\Nimp}}{1-\left(1-2\epsilon\right)^{\Nimp}}
%{1-\left(1-2\epsilon\right)^{\Nimp}}{1+\left(1-2\epsilon\right)^{\Nimp}},\\
%T^{\infty} &=& \frac{1}{2}\mymatrix{1}{1}{1}{1}.
%\end{eqnarray}
%\end{subequations}

We thus observe that the spin-flip scattering, when accumulated over many
events, leads to a complete mixing of spin, with the relative weightings
determined by $\epsilon_\sigma/(\epsilon_++\epsilon_-)$. In the special case of
spin-independent scattering, $\epsilon_+=\epsilon_-$, all matrix elements in the
r.h.s.\ of \eqref{eq:toy T spin-dep limit} are equal and hence, for a given
incoming spin direction, the transmission is mixed equally between both spin
directions.

The transmission $T^\infty$ reproduces qualitatively the behaviour of the
\emph{fractional} transmission from Section \ref{sub:spin-dep}, \ie
$T_{\sigma'\sigma}/(T_{\sigma\sigma}+T_{-\sigma\sigma})$. Obviously, such a
simplified picture does not reproduce the actual results on a quantitative
level. However, the qualitative resemblance suggests that it captures an
essential feature of transmission through a disordered domain wall, namely, a
spin-mixing effect which accumulates (and saturates) over successive scattering
events. 

\bibliography{bibliography}

\end{document}